\documentclass[aoas,reqno,preprint]{imsart}
\RequirePackage[numbers]{natbib}
\RequirePackage[OT1]{fontenc}
\RequirePackage{amsthm,amsmath}
\RequirePackage[colorlinks,citecolor=blue,urlcolor=blue]{hyperref}
\RequirePackage{multirow,bigdelim}
\usepackage{bold-extra}
\usepackage{ctable}
\usepackage{sparklines}
\usepackage{bm}
\usepackage{graphicx} 
\usepackage{import}
\usepackage{float}
\usepackage{bbold}
\usepackage{placeins}
\usepackage{mathtools}
\usepackage{chngcntr}
\usepackage{ulem}
\counterwithin{table}{section}
\usepackage{afterpage}
\usepackage{pdflscape,array,booktabs}
\usepackage{lipsum}
\usepackage{rotating}
\usepackage{flafter}
\usepackage[toc,page]{appendix}
\usepackage{footnote}
\usepackage{threeparttable}
\usepackage{tablefootnote}
\usepackage{soul}

\graphicspath{{02b_figures/}}

\startlocaldefs
\numberwithin{equation}{section}
\theoremstyle{plain}

\endlocaldefs

\newcommand{\ssnum}[1]{\textsuperscript{\citenum{#1}}}
\newcommand\Mycite[1]{\citeauthor{#1}~(\citeyear{#1})}
\newcounter{tableeqn}[table]

\newcommand{\hlt}[1]{\textcolor{orange}{#1}}

\makesavenoteenv{tabular}

\begin{document}
\begin{frontmatter}
  \title{Modeling National Latent Socioeconomic Health and Examination of Policy Effects via Causal Inference}

  \begin{aug}
    \author{\fnms{F.~Swen} \snm{Kuh}\thanksref{m1}\ead[label=e1]{swen.kuh@anu.edu.au}},
    \author{\fnms{Grace S.} \snm{Chiu}\thanksref{m4,m2,m3,m1}\ead[label=e2]{gschiu@vims.edu}}
    \and
    \author{\fnms{Anton H.} \snm{Westveld}\thanksref{m1}\ead[label=e3]{anton.westveld@anu.edu.au}}

    \affiliation{The Australian National University\thanksmark{m1}, William \& Mary's Virginia Institute of Marine Science\thanksmark{m4}, University of Washington\thanksmark{m2}, and University of Waterloo\thanksmark{m3}}

    \address{
      \printead{e1}\\
      \phantom{E-mail:\ }\printead*{e2} \\
      \phantom{E-mail:\ }\printead*{e3} }
  \end{aug}

\begin{abstract}
\[Abstract\]
This research develops a socioeconomic health index for nations through a model-based approach which incorporates spatial dependence and examines the impact of a policy through a causal modeling framework. As the gross domestic product (GDP) has been regarded as a dated measure and tool for benchmarking a nation's economic performance, there has been a growing consensus for an alternative measure---such as a composite `wellbeing' index---to holistically capture a country's socioeconomic health performance. Many conventional ways of constructing wellbeing/health indices involve combining different observable metrics, such as life expectancy and education level, to form an index. However, health is inherently latent with metrics actually being observable indicators of health. In contrast to the GDP or other conventional health indices, our approach provides a holistic quantification of the overall `health' of a nation. We build upon the latent health factor index (LHFI) approach that has been used to assess the unobservable ecological/ecosystem health. This framework integratively models the relationship between metrics, the latent health, and the covariates that drive the notion of health. In this paper, the LHFI structure is integrated with spatial modeling and statistical causal modeling, so as to evaluate the impact of a policy variable (mandatory maternity leave days) on a nation's socioeconomic health, while formally accounting for spatial dependency among the nations. We apply our model to countries around the world using data on various metrics and potential covariates pertaining to different aspects of societal health. The approach is structured in a Bayesian hierarchical framework and results are obtained by Markov chain Monte Carlo techniques.
\end{abstract}


\end{frontmatter}

\section{Introduction}
The gross domestic product (GDP) has been conventionally used as a measure when benchmarking different countries' growth and production. However, the commonly used GDP arguably only captures one aspect/perspective---the economic performance of a country---rather than a country's \textit{overall} performance and wellbeing. Consequently, many ongoing discussions and much effort have been made to find an alternative `wellbeing' indicator as a holistic measure of a country's socioeconomic health [\Mycite{conceiccao2008}]. Such wellbeing indices are useful for governments and organizations to benchmark a country's overall performance (other than solely economic) and help policy makers form evidence-based decisions. Despite that, there are issues with existing methods that attempt to quantify this health/wellbeing feature, for instance, combining multiple sources of subjectivity and arbitrarily turning them into a single score, yet without rigorously quantifying the uncertainties around the score [\Mycite{NEF2018}; \Mycite{OECD18}; \Mycite{UN2018}], or measuring a country's wellbeing using a chosen proxy variable such as the life satisfaction score [\Mycite{sachs2018}], which is not a direct measurement of the variable of interest. Health and wellbeing are increasingly being accepted as multidimensional concepts that often involve multiple subjective and objective measures on the macro- and micro-levels [\Mycite{mcgillivray2006}; \Mycite{yang2018}]. We recognize that the concept of wellbeing is inevitably subjective and we focus on reducing the subjectivity on the quantifiable measures through statistical inference of the country's socioeconomic health as a model parameter.

This paper proposes a hierarchical, latent variable framework to simultaneously model each country's health as a latent parameter, account for spatial correlation among countries, and evaluate the causal impact of a policy variable on the \textit{latent} health. This new methodology contributes to the aforementioned effort towards a holistic approach by addressing the subjectivity and uncertainty propagation through a single statistical inferential framework. We adapt the latent health factor index (LHFI) method [\Mycite{chiu2011}; \Mycite{chiu2013}] to quantify the country's `health' \textbf{$H$} as a latent parameter. Our work builds on the concept of assessing the underlying ecosystem health in \Mycite{chiu2011} and \Mycite{chiu2013} as unobservable and latent, to assessing societal health for countries. 

Note that the approach to measuring latent traits is not unique, as the idea appears in item response theory (IRT) in the psychometrics literature [\Mycite{rabe2004}]. Other examples include the quantification of the position of political actors on a political spectrum [\Mycite{jackman2001}; \Mycite{martin2002}], constructing measures of nations' underlying democracy [\Mycite{treier2008}], and assessing ecological/ecosystem health [\Mycite{chiu2011}; \Mycite{chiu2013}]. \Mycite{rijpma2016} model the wellbeing of countries also as a latent variable, similar to the special-case LHFI model that regresses health indicators on $H$ alone. In contrast, the general LHFI model further regresses $H$ on covariates that are chosen due to their perceived explanatory nature to health. In this paper, our holistic framework further incorporates spatial and causal modeling structure into the LHFI framework. 

In applying our work, we quantify the latent health of the countries using data collected at the national-level. Observable variables (e.g. gross national income (GNI) per capita, life expectancy, mean years of schooling, etc.) are treated as either \textit{indicators} or \textit{drivers/covariates} of a country's underlying health condition as opposed to \textit{measures} of health. We use `health' and `wellbeing' interchangably to capture the notion of a country's socioeconomic performance from the social, political, economic and environmental perspectives simultaneously. For the rest of the paper we will continue to refer to this holistic notion as (latent) health when referring to \textit{both} the model parameter and the concept of wellbeing. As national-level variables tend to be spatially dependent [\Mycite{ward2018}], we incorporate a spatial modeling structure into the LHFI framework to formally model this dependency among the countries. 

In addition to the quantification of the latent health of countries, the incorporation of causal modeling into our framework enables further insight into the effect of a policy variable on the health of a country. Propensity score adjustment for reducing confounding bias in observational studies has been used widely in the literature since the seminal paper by \Mycite{rosenbaum1983}. Subsequently, there have been ample discussions [\Mycite{an2010}; \Mycite{kaplan2012}; \Mycite{zigler2013}; \Mycite{zigler2014}] on modeling the uncertainty associated with the inference of the propensity score, as reflected by \Mycite{mccandless2009} who model the uncertainty under a Bayesian framework to evaluate the impact of statin therapy on mortality of myocardial infarction patients. We extend this idea to the context of evaluating the impact of a policy `treatment' variable (in our case, mandatory maternity leave (MML) days) on a country's health. Including this notion of `policy treatment' in our model allows a model-based assessment of the effect of a policy variable on the (latent) health of a country, in the context of counterfactuals.

To elaborate on the above elements, our paper is laid out as follows. In the next section, we briefly review the methodologies used to construct some of the existing socioeconomic health indices. Section \ref{sec:method} discusses the methodology and building blocks we employ to construct our latent socioeconomic health for nations. In Section \ref{sec:data}, we introduce the countries' data, propose a foundational framework (using the building blocks discussed in Section \ref{sec:method}) then an extended version to be applied to the data, and highlight some of the results from our models. In Section \ref{sec:discussion}, we revisit the data by providing an in-depth discussion of the specifics of the data and model structure we have used. Finally, we review the limitations of our work and conclude the paper by discussing some potential future work in Section \ref{sec:future}. Supplementary materials are included in the appendix.

\section{Review on existing indices}
\subsection{Global and regional indices}
There is an increasing awareness that the GDP has been inappropriately used as a broader benchmark measure for overall welfare among countries [\Mycite{ida2014}]. Several methods have been proposed as an alternative measure to the GDP, but existing approaches have used variables such as the life evaluation score or `happiness' as a proxy measure of a country's health (or subjective wellbeing) [\Mycite{sachs2018}; \Mycite{conceiccao2008}]. This is also problematic, as a country's health is a multidimensional concept as aforementioned. We review five such alternative indices in Table \ref{tab:indices}. The background and components contributing to these five and other indices have been discussed by \Mycite{hashimoto2018} and \Mycite{ida2014}, but here we focus on the statistical methodology being used. Note that 2 out of these 5 indices assume equal weighting of pre-specified variables that contribute to a country's health. There appears to be little justification that the concept of health is represented by equal parts of a wide variety of variables, apart from convenience.

\begin{table}[h!] \label{tab:indices}
  \caption{Selection of existing indices and methodology used}
  \begin{tabular}{|p{15em}|p{20em}|}
    \hline
    \textsc{\bfseries Index} & \bfseries{\scshape Statistical Methodology} \\
    \hline
    \textbf{United Nations Human Development Index (HDI)}\ssnum{UN2018}& Arithmetic means of different variables are computed, then a geometric mean of the arithmetic means is computed to form the HDI \\
    \hline
    \textbf{World Happiness Report}\ssnum{sachs2018}                                                             & Pooled ordinary least squares regression (from econometrics) of the national average response to the survey question of life evaluations on 6 categories of variables hypothesized as underlying determinants of the nation's `happiness score'\\
    \hline
    \textbf{Social Progress Index (SPI)}\ssnum{stern2018}                                                              & First, a principal component analysis (PCA) is used to determine the weighting of indicators within each component, and the weights and indicators are multiplied to obtain component scores. Next, component scores are transformed onto a scale of 0-100, an arithmetic mean is computed for each dimension, and another arithmetic mean is computed to obtain the final SPI \\
    \hline
    \textbf{Happy Planet Index (HPI)}\ssnum{NEF2018} & The variables `experienced wellbeing' and `life expectancy' are multiplied, then divided by `ecological footprint'; scaling constants are used to map the final HPI to range from 0-100  \\
    \hline
    \textbf{Organisation for Economic Co-operation and Development (OECD) Better Life Index (BLI)}\ssnum{OECD18} & OECD BLI website user-specified weights are assigned to each topic (e.g. education, income, etc.), and up to four indicators which constitute each topic are assigned equal weights to form the final BLI \\
    \hline
  \end{tabular}
\end{table}

\section{Methodology}\label{sec:method}
To quantify the \textit{latent socioeconomic health} and its uncertainty in a policy-specific context, we integrate two novel approaches---the latent health factor index (LHFI) [\Mycite{chiu2011}] and the Bayesian propensity score analysis (BPSA) [\Mycite{mccandless2009}]---along with spatial modeling to account for spatial dependence among countries. We are interested in these two methods as the former describes health as \textit{latent}, i.e. something that is not directly measurable, while the latter allows us to examine the effect of policy prescription and to quantify the effect on the nation's overall socioeconomic health.

\subsection{Latent health}\label{sec:LHFI}
\begin{figure}[H]
\caption{(adapted from \Mycite{chiu2013})}
\begin{center}
  \includegraphics[width=0.7\textwidth]{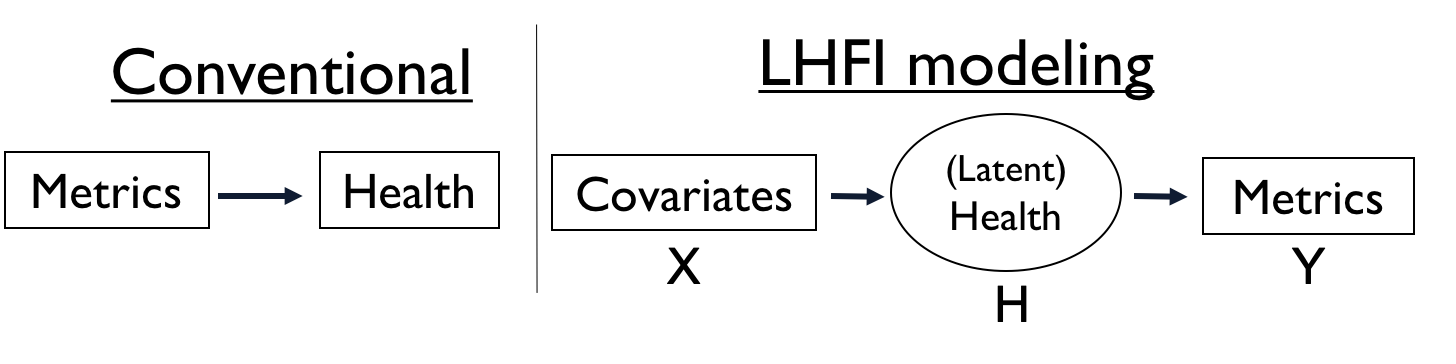}\label{fig:LHFI}
\end{center}
\end{figure}

As an analogy to a country's latent health, the underlying health conditions of a person who is deemed healthy cannot be directly compared to those conditions of another person. It is the measurable variables such as height, weight or calorie intake of a person that can be compared. Similarly, for a country, there is no single directly observable quantity that can represent ``how well a country is doing". Thus, the health of a country is a notion that we wish to evaluate comprehensively and holistically. For instance, we may argue that variables like GDP, life expectancy, and infant mortality rate can each coarsely inform us on some aspect of the state in which a country's health is, but not its \textit{overall} health. The LHFI framework unifies multiple aspects of health by modeling the underlying condition that we wish to assess as a \textit{latent} parameter (not directly measurable), but it is dependent on different measurables that are either drivers of health (covariates), or indicators of health (metrics). A schematic representation of the LHFI framework is shown in Figure \ref{fig:LHFI}.
\vspace{-1em}
\vspace{-1.5em}
The LHFI structure employed to model health as a latent variable for our specific context is a type of mixed model [\Mycite{rabe2004}], where nation-specific health is a  random effect. We formulate our model as a Bayesian hierarchical mixed model, as it is noted in \Mycite{gelman2007} as the most direct approach to handle latent structures. 

To avoid parameter identifiability problems that are prominent in item response models (a type of generalized linear mixed model) as discussed in \Mycite{martin2002}, we used a semi-informative prior on one pre-selected country's health parameter, $H_{anc}$, to anchor our latent score's scale. A more in-depth discussion of the anchoring approach can be found in Section \ref{ssec:ident}. 

Note that a country's metrics are multivariate in nature. Thus, in our hierarchical model, we use a multivariate normal distribution on the first level in the hierarchy, i.e. the metric-level (Y-level) (equation (\ref{eq:MVNy})). Therefore, our base LHFI model (excluding BPSA and spatial elements) with an `$H$-anchor' takes on the form below: 

\vspace{-1.2em}
\begin{align}
  \bm{y_i} | \bm{a}, H_i, \Sigma_Y                    & \stackrel{ind.}{\sim} \text{MVN}(\bm{a}H_i, \Sigma_{Y}) \label{eq:MVNy} \\
  \bm{H}_{i\neq anc} | \bm{\beta}, \bm{X}, \sigma^2_H & \sim \text{MVN}(\bm{\mu}_{i\neq anc}, \Sigma_H)  \label{eq:MVNh}        \\
  {H_{anc} }                                          & \sim \text{N}(-2, 0.1) \label{eq:Hanc} \\
  \text{where}  \ \ \
  \bm{\mu}_{i\neq anc}                                & = [\bm{X}\boldsymbol{\beta}]_{i\neq anc} \notag  \\
  \Sigma_H                                            & = \sigma^2_H{\boldsymbol {I}}_{(N-1)} \notag 
\end{align}
and where MVN and N denote the multivariate and univariate normal distributions, respectively. Normality is assumed due to the nature of our metric variables (see Section \ref{sec:data}).
 
At the Y-level, we let $\bm{y_i} = (y_{i1}, \ldots, y_{iP})^T$ be a $P \times \text{1}$ vector for the $i$th country's metrics for $i = 1, \ldots, N$; $\bm{a} = (a_1, \ldots, a_P)^T$ be the $ P \times 1$ vector for the `loadings' of any country's health on its metrics; and $\Sigma_Y$ be the $P \times P$ covariance matrix for the metrics. \

We refer to equation (\ref{eq:MVNh}) as the health-level ($H$-level), where $\bm{H}_{i\neq anc}$ is an $(N - 1) \times \text{1}$ vector of \textit{latent} health for all $N$ countries \textit{except} for the chosen anchor country; $\bm{\mu}_{i\neq anc}$ is the corresponding linear regression mean evaluated using $\bm{X}$ and $\bm{\beta}$;
$\bm{X} = (\mathbf{1}, X_1, \ldots, X_K)$ is an $N \times (K+1)$ matrix with $K$ different covariates and $\mathbf{1}$ is an $N \times 1$ vector of ones; $\bm{\beta} = (\beta_0, \beta_1, \ldots, \beta_K)^T$ is a $(K+1) \times 1$ vector including the intercept and associated slope coefficient corresponding to the \textit{k}th covariate; $\sigma^2_H$ is the common variance for $H_i$ for all $i$ excluding $H_{anc}$; ${\boldsymbol {I}_{(N - 1)}}$ is the $(N - 1) \times (N - 1)$ identity matrix; finally, $H_{anc}$ for the chosen anchor country follows a semi-informative normal prior distribution with fixed mean and variance (equation (\ref{eq:Hanc}); see Section \ref{ssec:ident} for justification). \

We include the intercept $\beta_0$ at the $H$-level rather than the $Y$-level because we have standardized each metric to have mean zero and unit variance, due to the vastly different scales among metrics. A discussion of data transformation is found in Section \ref{ssec:missing}.

\subsection{Spatial modeling}\label{sec:spatial}
Macro-level variables of countries are expected to be spatially correlated [\Mycite{ward2018}], as countries that are close together in regions (e.g. Europe, North America and Central Asia) tend to be more similar in terms of a cultural, economic, social or political context. This suggests that \textit{latent health} may also be spatially dependent. In order to assess the need for spatial modeling in our framework, we fit the base LHFI model using equations (\ref{eq:MVNy} - \ref{eq:Hanc}) and examined its residuals. The residuals $\epsilon_H$ are calculated using

$$\bm{\widehat{\epsilon_H}}^* = \bm{\widehat{H}} - \bm{X\widehat{\beta}} $$
where the `hat' ($\ \widehat{} \ $) values are the posterior medians of the parameters based on the Markov chain Monte Carlo (MCMC) samples. \

\begin{figure}[!ht]
  \caption{Residuals $\epsilon_H^*$ on the world map}
  \includegraphics[width=\textwidth]{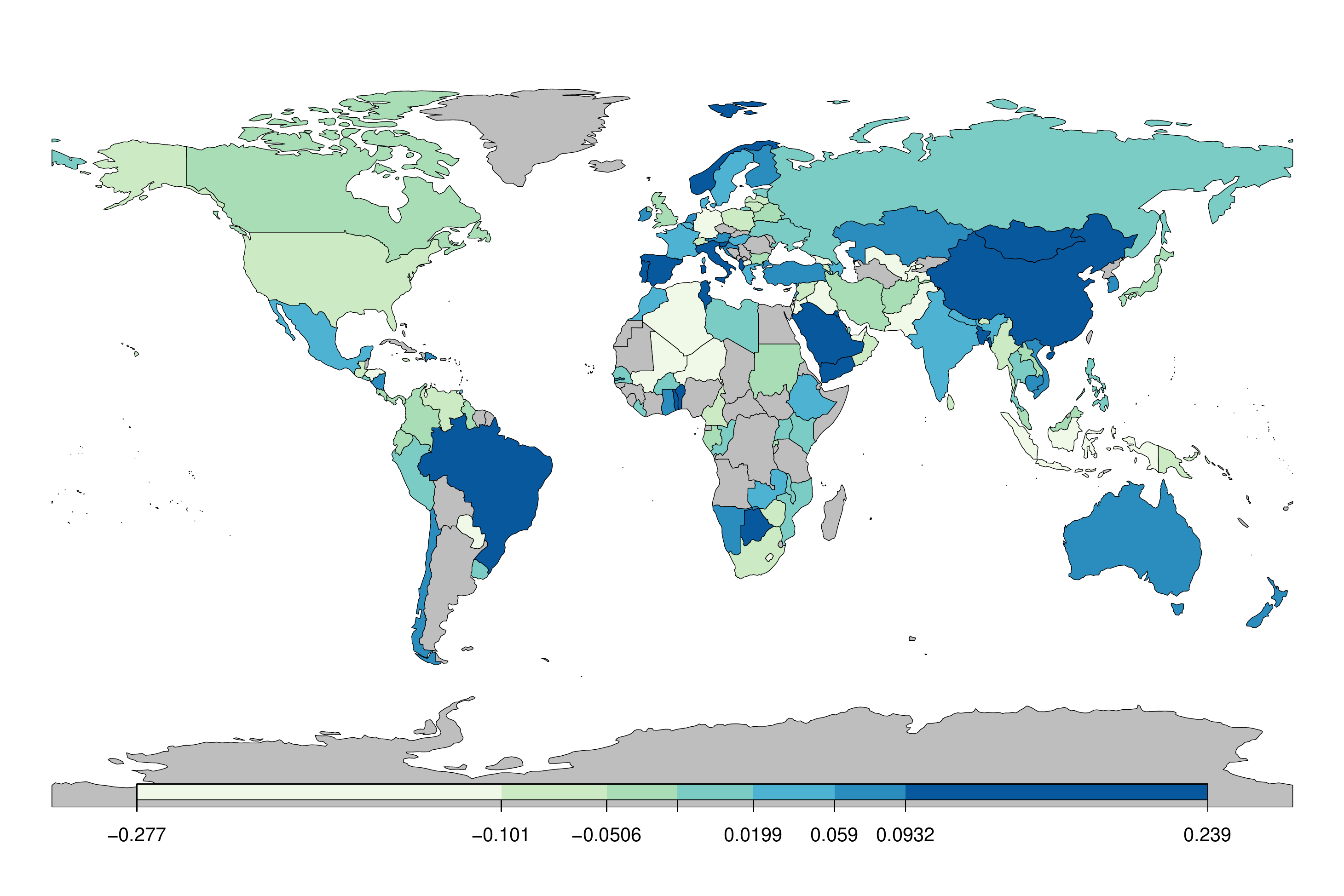}\label{fig:Hres}
  \vspace{-3em}
\end{figure} \nocite{south11}

Figure \ref{fig:Hres} presents the estimated residuals from an initial LHFI model fit on the world map. It is apparent that countries that are geographically close together in regions such as North America, Western Europe, Sub-Saharan Africa and South East Asia have posterior residuals that are either similarly under- or over-estimated by our model. To accommodate this, we incorporate spatial dependency among our residuals on the health-level, and modify equation (\ref{eq:MVNh}) to account for spatial dependence in its residuals, resulting in a spatial LHFI model:

\begin{align}
\bm{y_i} | \bm{a}, H_i, \Sigma_Y &\stackrel{ind.}{\sim} \text{MVN}(\bm{a}H_i, \Sigma_{Y}) \label{eq1} \\ 
  {\bm{H}_{i\neq anc} | \bm{\beta}, \bm{X}, \Sigma_H} & \sim \text{MVN}(\bm{\mu}_{i\neq anc} , \Sigma_H) \label{eq:spa0} \\
  {H_{anc} }                                          & \sim \text{N}(-2, 0.1)                                           \\
  \text{where} \ \ \
  \bm{\mu}_{i\neq anc}                                & = [\mathbf{X}\boldsymbol{\beta}]_{i\neq anc}                     \\
  \Sigma_H                                            & = \sigma_H^2 \mathbf{R}(d,\phi) \label{eq:spa1}                  \\
  \mathbf{R}(d,\phi)                                  & = \begin{bmatrix}
    1         & \rho_{nm} & \cdots    & \rho_{nm} \\
    \rho_{mn} & 1         & \ddots    & \vdots    \\
    \vdots    & \ddots    & \ddots    & \rho_{nm} \\
    \rho_{mn} & \cdots    & \rho_{mn} & 1         \\
  \end{bmatrix} \label{eq:spa2}                      \\
  \rho_{nm}                                           & = exp(-d_{nm}/\phi) \label{eq:spa3}                       
\end{align}

where $\Sigma_H$ denotes the $(N-1) \times (N - 1)$ spatial covariance matrix for \textit{health}; $\rho_{nm}$ is the correlation parameter between countries $n$ and $m$, which is a function of  $d_{nm}$ (the distance between two countries) and $\phi$ (the `range' or inverse rate of decay parameter).

The covariance function we employ in equations (\ref{eq:spa1} - \ref{eq:spa3}) is a special case of the Mat\'ern class of spatial covariance functions, for modeling the dependence between spatial observations [\Mycite{gelfand2010}]. For instance, a large value of $\rho$ suggests that countries that are relatively far from one another are still moderately correlated [\Mycite{hoeting2006}]. Note that geographical distance measures on a global scale have always been a contentious issue [\Mycite{ward2018}]. We discuss some of the possible extensions to the spatial component in our framework in Section \ref{ssec:spatial}.

\subsection{Causal inference}\label{sec:BPSA}
In addition to quantifying our latent socioeconomic health and its uncertainty, we seek to integrate causal modeling into our framework to provide insight into the effect of a `policy treatment' variable on the health of a country.

Two contending schools of thought dominate the causal inference literature --- namely, ``Pearl's causal diagram'' [\Mycite{pearl2009}] and ``Rubin's causal model'' [\Mycite{imbens2015}]. Both attempt to establish causal effects from observational studies, which was previously considered impossible because such studies are not randomized controlled trials [\Mycite{imbens2015}; \Mycite{hernan2018}]. In our current work, we consider the propensity score (PS), which is one of the methods in Rubin's approach to estimate the average treatment effect. Among causal inference methods for non-experimental data, propensity score analysis (stratification, matching and covariate adjustment) has been widely used to address selection bias [\Mycite{imbens2015}]. There has since been research that considers the uncertainty in the propensity scores [\Mycite{mccandless2009}; \Mycite{an2010}], although incorporating the outcome variable at the stage where the inference of PS is conducted may be contentious [\Mycite{kaplan2012}; \Mycite{zigler2013}; \Mycite{zigler2016}]. For this reason, our framework on including the PS is built on \Mycite{zigler2013}'s work, which uses Bayesian posterior-predictive methods to separate the design and analysis stage in order to `cut the feedback' (i.e. to ensure that the inference of the PS does not depend on the outcome variable) [\Mycite{mccandless2010}; \Mycite{zigler2013}; \Mycite{zigler2014}].

There are three main assumptions in Rubin's approach of causal modeling, namely, the i) stable unit treatment value assumption (SUTVA) which stipulates no interference between units [\Mycite{rubin2005}]; ii) strongly ignorable treatment assignment which stipulates no unmeasured confounders [\Mycite{rosenbaum1983}]; and iii) consistency, where the potential outcome of the treatment must correspond to the observed response when the treatment variable is set to the observed `exposure' level [\Mycite{cole2009}]. 

However, incorporating causal modeling in a spatial setting potentially violates the no interference assumption (SUTVA) as discussed at length in \Mycite{keele2015} and \Mycite{noreen18}. In investigating the effect of convenience voting and voter turnout, \Mycite{keele2015} are concerned about interference and spillover effects -- the units (individuals) may be influenced due to proximity of geographical regions (or influenced at the workplace or by their social network, etc.); hence, assuming spatial dependence may violate the SUTVA assumption. 

In our paper, the causal question of interest is the effect of a national policy variable on a country's health, with the unit of interest being at the national-level rather than at the individual-level. Two obvious scenarios of interference  and spillover in our case may be i) due to individuals immigrating or emigrating and in their newly adopted country, either influencing policy makers or affecting the \textit{health} of the country (e.g. a Canadian mother whose wellbeing benefited from the Canadian federal maternity leave policy emigrating to the United States, which does not have federal maternity leave, might improve the \textit{health} of the United States.); ii) policy makers influenced by their international social networks. However, it should be reasonable to assume the effects of individuals' international migration on a nation's maternity leave policy and \textit{health} to be minimal. Additionally, it should be reasonable to assume that federal policy making --- especially regarding maternity leave --- is a collective domestic effort and generally conducted with minimal foreign interference. Finally, as shown in \Mycite{schutte2014}, when there is little overlap in the units, consistent treatment effect can still be valid. Note that we dichotomized the continuous policy treatment variable into a binary variable, so that countries are defined to be either below or above the median of this policy variable, also with no overlap. (How best to utilize a continuous treatment variable in the context of PS is discussed in Section \ref{sec:future}.) These arguments suggest that it is reasonable for us to assume that SUTVA holds.

Moreover, we select our covariates based on structural variables or a country's existing infrastructure. By conditioning on the covariates through propensity scores, we assume that there are no unmeasured important confounders. 

As such, we proceed with the PS framework while assuming these three assumptions hold. To incorporate causal modeling into our spatial LHFI model, we introduce the policy treatment variable $T$ and its propensity scores $z(X_i, \bm{\gamma})$ to our health-level through its mean:

\vspace{-2em}
\begin{figure}
\begin{center}
\caption{Extension of the LHFI framework with causal modeling}
  \includegraphics[width=0.7\textwidth]{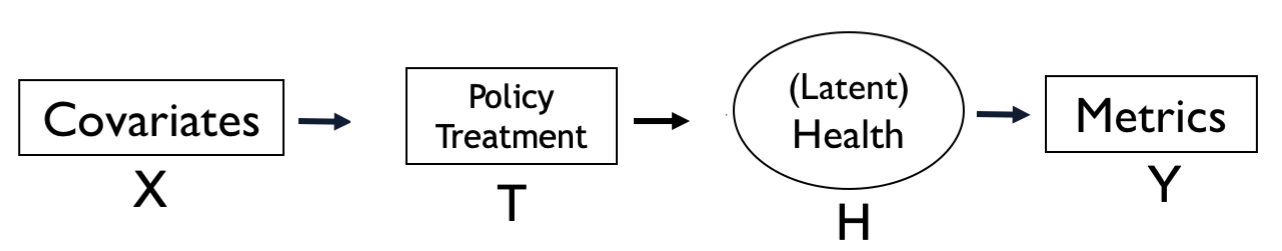} \\
\end{center}
\vspace{-5em}
\end{figure}

\begin{align}
\bm{y_i} | \bm{a}, H_i, \Sigma_Y &\stackrel{ind.}\sim \text{MVN}(\bm{a}H_i, \Sigma_{Y}) \label{eq:B1} \\ 
  \bm{H_{i\neq anc}} |  \bm{T}, \bm{\beta}, \bm{X}, \bm{\gamma}, \bm{\xi}, \Sigma_H & \sim \text{MVN}(\bm{\mu}_{i\neq anc}, \Sigma_H)  \label{eq2}                 \\
  \text{logit}[\text{P(} T_i = 1 | X_i)] & =  X_i\bm{\gamma}  \label{eqLR}                                                 \\ \vspace{2em}
  {H_{anc} }                                                                                        & \sim \text{N}(-2, 0.1)                                                       \\
  \text{where}  \hspace{3em}
  \bm{\mu}_{i\neq anc}                                                                              & = [\bm{T\beta }   +  \bm{g(z(X, \bm{\gamma}))\bm{\xi}}]_{i\neq anc} \label{eq:PS_mean} \\
  \Sigma_H                                                                                          & = \sigma_H^2 \mathbf{R}(d,\phi) \label{eq:3.32}                              \\
  \mathbf{R}(d,\phi)                                  & = \begin{bmatrix}
    1         & \rho_{nm} & \cdots    & \rho_{nm} \\
    \rho_{mn} & 1         & \ddots    & \vdots    \\
    \vdots    & \ddots    & \ddots    & \rho_{nm} \\
    \rho_{mn} & \cdots    & \rho_{mn} & 1         \\
  \end{bmatrix} \label{eq:spa4}                      \\
  \rho_{nm}                                           & = exp(-d_{nm}/\phi) \label{eq:spa5}
\end{align}
\vspace{-1.8em}

The parameters are similar to the spatial LHFI model (eq. \ref{eq:spa0}--\ref{eq:spa3}) in the previous section except for $\bm{\mu}_{i \neq anc}$, the health-level mean for all non-anchor countries. Also, there is the propensity score $z(\cdot)$ through the indicator function $g(\cdot)$ and its associated coefficient $\bm{\xi}$ in equation (\ref{eq:PS_mean}). Note that $\bm{T} = (\mathbf{1}, T)$ is an $N \times 2$ matrix where  $\mathbf{1}$ has length $N$, and vector $T$ of length $N$ is the policy treatment variable of interest; $\bm{\beta} = (\beta_0, \beta_1)^T$ and $\beta_1$ now represents the \textit{average policy treatment effect} on a country's \textit{health}; $\bm{X} = (X_1, \ldots, X_K)^T$ is now an $N \times K$ matrix of covariates and $\bm{\gamma}$ is a $K \times 1$ vector of logistic regression coefficients; and $X_i$ is the $i$th row of $\bm{X}$.

Similar to the set up in \Mycite{mccandless2009}, we let $g(z(X_i, \bm{\gamma}))$ be a 2 x 1 vector of indicator variables that models the membership of countries within one of the three subclasses:

\vspace{-1.8em}
\begin{align*}
 z(X_i, \bm{\gamma})   &= \text{expit}(X_i\bm{\gamma}) \ \ (\text{Propensity score}) \\
 g(z(X_i, \bm{\gamma}))^T & =
  \left\{
  \begin{array} {l}
    {[}0,0{]} \qquad \text{if} \ \  0<Z_i<q_1        \\
    {[}1,0{]}\qquad \text{if} \ \  q_1\leq Z_i < q_2 \\
    {[}0,1{]}\qquad \text{if} \ \  q_2\leq Z_i < 1
  \end{array}
  \right.
\end{align*}

Note that the propensity score $Z_i = \text{P}(T_i=1|X_i) = z(X_i, \bm{\gamma})$ is interpreted as the probability (under a logistic regression model) that $T_i = 1$ given $X_i$ and $\bm{\gamma}$. If $\bm{\gamma}$ is known, the inferred propensity score is computed using $z(X_i, \bm{\gamma}) = \text{expit}(X_i\bm{\gamma})$ where $\text{expit}(a) = (1+ \text{exp}(a))^{-1}$. Our vectors $g(\cdot)$ are of length 2 as we have introduced an intercept term $\beta_0$ at the $H$-level. Hence, $\bm{\xi} = (\xi_1, \xi_2)^T$ is the corresponding regression coefficients for the indicator function $g(\cdot)$ characterized by the intervals [$q_1$, $q_2$), [$q_2$, 1]. The varying knot values $q_1$ and $q_2$ for defining the quantiles of $Z$ are determined in each MCMC iteration, by allocating roughly one-third of the countries to each tertile of the propensity scores. While \Mycite{rosenbaum1983} suggest stratifying on quintiles of the propensity score for removing 90 per cent of the bias from measured confounders, this stratification of propensity scores into three subclasses is decided based on the relatively small sample size in our data, and after assessing the balance in treatment and control groups with respect to the covariates (see discussion in Section \ref{ssec:PS}).

\section{Latent health for the World}\label{sec:data}
\subsection{Data}\label{ssec:data}
We collated our data from the years 2010--2015 from publicly available databases. However, in this paper, we focus on the results for 2015, being the most recent year available at the time of data collation. We have consolidated 15 metrics, 1 treatment variable and 4 covariates shown in Table \ref{tab:variables}. Most of the metrics and covariates employed in our models are taken from the data section in the United Nations Human Development Report, which is sourced from various organizations and the World Bank database. Specifically, the POLITY variable is sourced from the Polity IV project [\Mycite{polity}], and Corruption Perception Index from the Transparency International website [\Mycite{CPI}]. Other relevant variables (e.g. literacy rate among adults in the country) were not included in our model due to a substantial amount of missing data. \

\begin{table}[htb!]
\label{tab:variables}
\caption{List of variables X, T and Y}
\begin{tabular}{llc}
\toprule
\textbf{X, Covariates}& \textbf{T, Treatment variable} \\
\midrule
Forest area & Federally mandated maternity leave days \\
Access to electricity, rural & \\
Mean years of schooling & \\
Population, total & \\
\midrule
\multicolumn{2}{c}{	\textbf{Y, Metrics}}  \ML
Education index & Population density \\
Popn., urban (\% of total) & Popn., ages 65 and older \\
Employment to popn. ratio, (\%) & Unemployment rate (\%) \\
Corruption Perception Index & Life expectancy \\
Infant mortality rate & Internet users (\% of popn.) \\
Renewable energy consumption (\%) & POLITY index \\
Gross National Income (GNI) \NN \ \ per capita (current international \$) &  \\
Prop. of parliamentary seats \NN \ \ held by women (\%) & \\ 
Popn. with at least some \NN \ \ secondary education (\% ages 25 and older) & \\
\bottomrule
\end{tabular}
\vspace{-1em}
\end{table}

As the covariates $X$ in this framework are regarded as the \textit{drivers} of a country's socioeconomic health, we chose them based on the country's resources and existing infrastructure (e.g. forest area). The $Y$'s are \textit{indicators} of health (e.g. education index) based on measures that we perceive as \textit{reflective} of a country's health. In particular, GNI as opposed to GDP was used as it is perceived as a more inclusive indicator of a country's wealth [\Mycite{klugman2011}]. These indicators, or metrics, have been \textit{a priori} transformed so that increasing values reflect better health; see Section \ref{ssec:missing} for additional details. For our policy treatment variable $T$, we have selected the federally mandated number of maternity leave (MML) days in a country. This variable was chosen due to its proposed benefits to individuals, the economy and society as a whole [\Mycite{chapman2008}]. However, the World Bank data source only has alternate years of data for this policy variable, and we had to informally impute the data for some of the OECD countries using data from the OECD website [\Mycite{OECD}]. A discussion of data imputation is found in Section \ref{ssec:missing}.

To utilize the treatment variable in a PS framework, we dichotomized it at the median, as discussed in Section \ref{sec:BPSA}. For 2015, the median of the treatment variable is 98 federally MML days and it  ranges from zero (3 countries, e.g. USA) to 410 (Bulgaria). Note that according to the World Bank, Sweden has zero MML days based on its definition. However, the OECD and other sources suggest that this may not be an accurate representation of their maternity policy. (Future iterations of our work will consider non-World Bank definitions. Also, in Section \ref{sec:future} we discuss our intention to employ the untransformed numerical MML variable in future work.) It is recognized that the selection of modeled variables is inevitably subjective but could be informed by the modeler's domain knowledge. As such, in this paper we focus on the methodology and its interpretation. 

We present results from two models, model A (spatial LHFI in Section \ref{ssec:modelA}) and model B (causal spatial LHFI in Section \ref{ssec:modelB}), respectively fitted to the 2015 data. For Bayesian inference, the full conditional distributions for $\phi$ and $\bm{\gamma}$ were sampled using a Metropolis-Hastings algorithm due to non-conjugate priors, but all other parameters were sampled using Gibbs sampling. For each of models A and B, we utilized roughly 200,000 post-burn-in MCMC samples from the posterior distribution. Standard diagnostics using the \texttt{coda} package in R and examining trace plots ensured that each parameter of the MCMC chain had reached its steady state. 

\subsection{Model A: Base LHFI + Spatial modeling}\label{ssec:modelA}
This spatial LHFI model corresponds to Section \ref{sec:spatial}. For $N$ = 125 countries, we regress $P$ = 15 metrics jointly on $H$, and $H$ is in turn regressed on $K$ = 5 variables in total -- including 4 covariates $X$, and 1 treatment variable $T$.

\subsection*{Model A: Priors}
We specify conjugate diffuse priors for most parameters. For each regression coefficient $a_j$ and $\beta_k$, we specify a normal prior distribution with mean 0 and variance 100. The covariance matrix $\Sigma_Y$ is given an inverse-Wishart prior with $P + 2$ degrees of freedom, and a diagonal scale matrix with diagonal values equal to 1; and $\sigma^2_H$ is assigned an inverse-Gamma prior with shape = 1 and scale = 0.1. The spatial correlation inverse decay rate $\phi$ is modeled with a log normal prior with a mean of 0.4 and a standard deviation of 2. Finally, a semi-informative normal prior is used on the anchor country's \textit{latent health} ($H_{anc}$). A further discussion of the identifiability/interpretability of parameters for the health-level of the model is in Section \ref{ssec:ident}.

\subsubsection{Model A: Ranking of countries according to latent health, $H$}\phantomsection{\hspace{0.5em}}
\vspace{-1.2em}
\begin{figure}[H]
\caption{Model A: Latent health for 125 countries in 2015 grouped by income group}
\includegraphics[width=\textwidth, height=0.58\textheight]{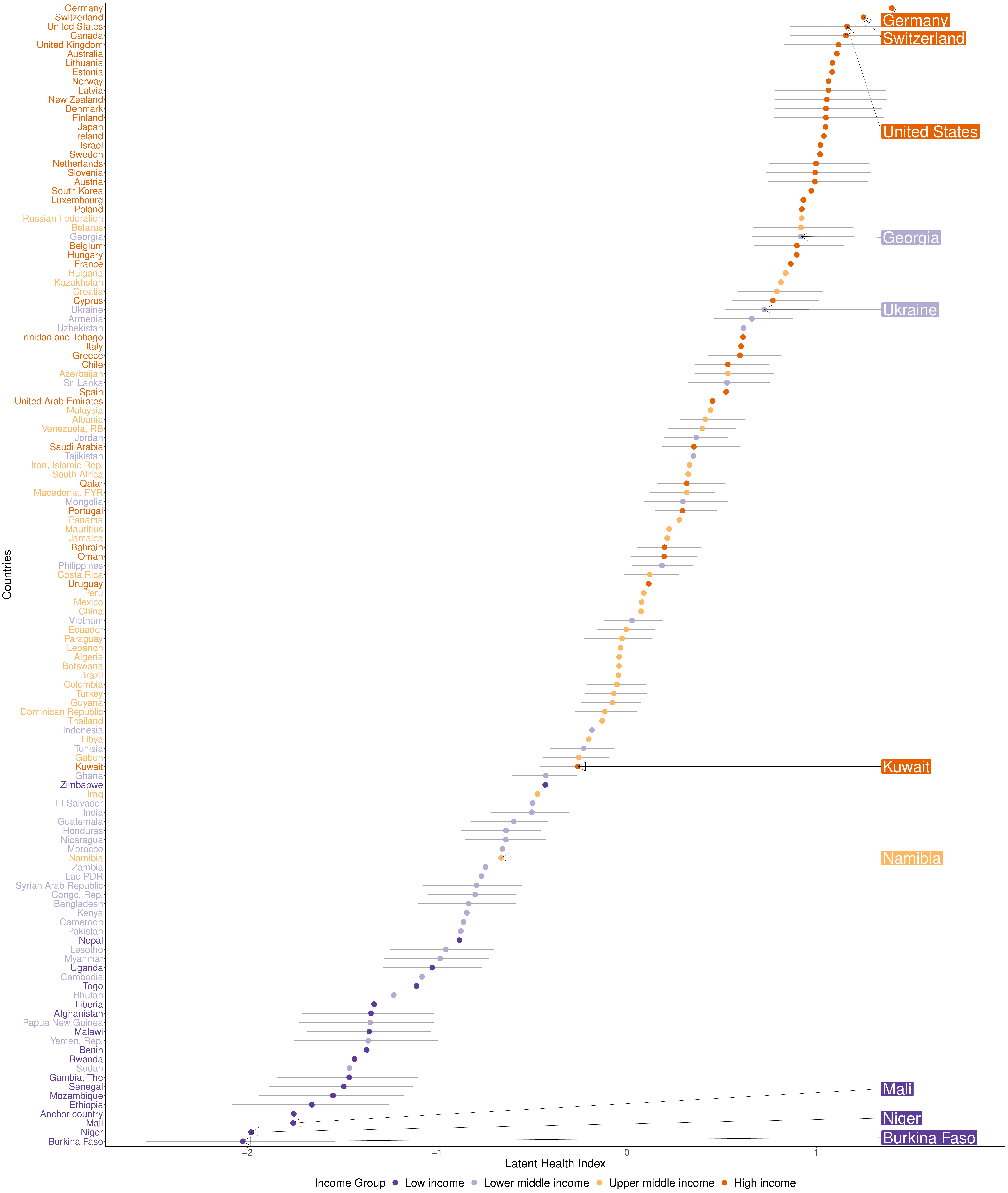}
\label{fig:HrankA}
\vspace{-1em}
\end{figure}

Figure \ref{fig:HrankA} shows the ranking of countries based on the posterior medians of the $H$'s (colored dots) along with their corresponding 95\% credible intervals (gray band). Through formal quantification of the uncertainty for our health parameter, the difference between countries is not polarized into developed/developing countries or rich/poor countries; the lack of polarization aligns with the findings in \Mycite{rosling2019}. 

Nevertheless, our color-coding according to the United Nations' designated income groups shows that the countries are generally ranked according to their income group. This suggests that the health of a country is highly correlated with the income group of the country. However, as will be discussed below, income is not the most important index to examine when considering the health of a country. Finally, the figure highlights some countries that are ranked highest, lowest, or differently than its income group. 

\subsubsection{Model A: Results}\label{ssec:resA}
Posterior summaries for some key parameters are shown in Table \ref{tab:param1}. Results for other parameters are tabulated in Appendix \ref{appendix:appB}.

\begin{table}[!htbp]
	\caption{Posterior summaries for selected model A parameters}
\begin{tabular}{ccccc}
\toprule
Parameter & Markov Chain & 2.5\% & Median (Mean) & 97.5\% \\
\midrule
$a_1$ & \includegraphics[width=0.1\textwidth, height=0.02\textheight]{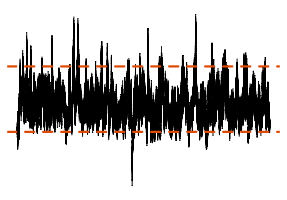} & \hlt{0.90} & 1.10 & \hlt{1.44}  \\
$a_8$ & 
\includegraphics[width=0.1\textwidth, height=0.02\textheight]{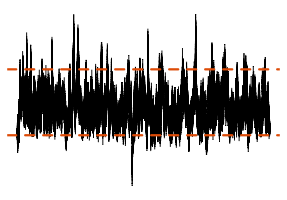} & \hlt{0.87} & 1.07 & \hlt{1.40} \\
$a_4$ & 
\includegraphics[width=0.1\textwidth, height=0.02\textheight]{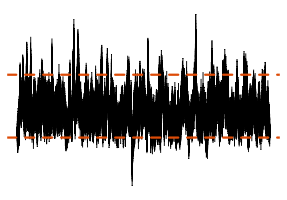} & \hlt{0.79} & 1.00 & \hlt{1.32} \\ 
$a_3$ & 
\includegraphics[width=0.1\textwidth, height=0.02\textheight]{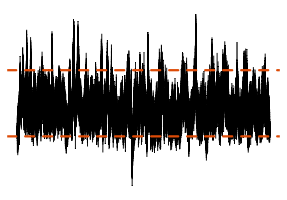} & \hlt{0.77} & 0.97 & \hlt{1.29} \\
$a_2$ & 
\includegraphics[width=0.1\textwidth, height=0.02\textheight]{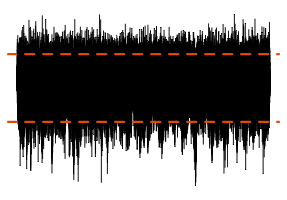} & \hlt{-0.49} & -0.27 & \hlt{-0.07} \\
$H_{28\phantom{,} \text{(Germany)}}$ & \includegraphics[width=0.1\textwidth, height=0.02\textheight]{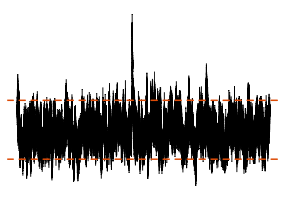} &  \hlt{1.04} & 1.40 & \hlt{1.77} \\
$H_{20\phantom{,} \text{(Switzerland)}}$ & \includegraphics[width=0.1\textwidth, height=0.02\textheight]{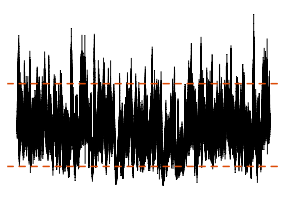} &  \hlt{0.93} & 1.26 & \hlt{1.58} \\
$H_{118\phantom{,} \text{(United States)}}$ & \includegraphics[width=0.1\textwidth, height=0.02\textheight]{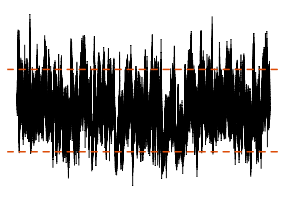} &  \hlt{0.86} & 1.17 & \hlt{1.49} \\
$H_{19\phantom{,} \text{(Canada)}}$ & \includegraphics[width=0.1\textwidth, height=0.02\textheight]{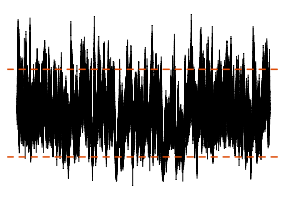} & \hlt{ 0.86} & 1.16 &  \hlt{1.47} \\ 
$\beta_1$  & \includegraphics[width=0.1\textwidth, height=0.02\textheight]{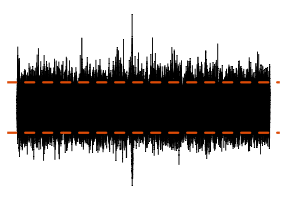}&  \hlt{ -0.04}  & 0.03 &  \hlt{ 0.10} \\ 
$\phi$ & \includegraphics[width=0.1\textwidth, height=0.02\textheight]{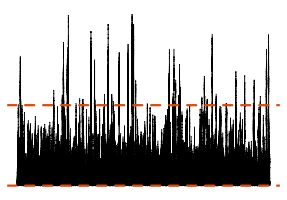} & \hlt{ 0.08} & 1.48 (3.50) &  \hlt{20.41} \\
\bottomrule
\end{tabular}
\label{tab:param1} 
\end{table}

\subsubsection*{Metric effects, $\bm{a}$} \label{ssec:metric}
Insights into the associated strength and direction of relationship between metrics and health are available from the inference about the health loadings, $a_j$. 

Table \ref{tab:param1} shows the results for the four loadings which have the highest positive impact, based on the medians of each of the marginal posterior distributions, and one example of loading that has a negative impact. In decreasing order of effect size, the corresponding positive metrics are: `\texttt{education index}' ($j$ = 1), `\texttt{proportion of population with at least secondary education}' ($j$ = 8), `\texttt{proportion of internet users}' ($j$ = 4), and `\texttt{GNI per capita}' ($j$ = 3). This suggests that, among the 15 metrics, education-themed variables receive the highest two loadings from the country's latent health. In contrast to conventional belief, the national wealth indicator `\texttt{GNI per capita}' is ranked fourth in terms of its positive association with a country's health. In fact, the posterior probability for health to have a bigger effect on `\texttt{education index}' than `\texttt{GNI per capita}' is 0.99. This suggests with rigor that a country's health is not solely reflected by a country's wealth, but other social factors as well. 

Interestingly, three metrics were found to have a negative relationship with health. 

\vspace{-1.5em}
\begin{figure}[H]
\caption{Plot of employment to population ratio vs. posterior median of latent health with a least squares regression fit }
\includegraphics[width=0.9\textwidth, height=0.35\textheight]{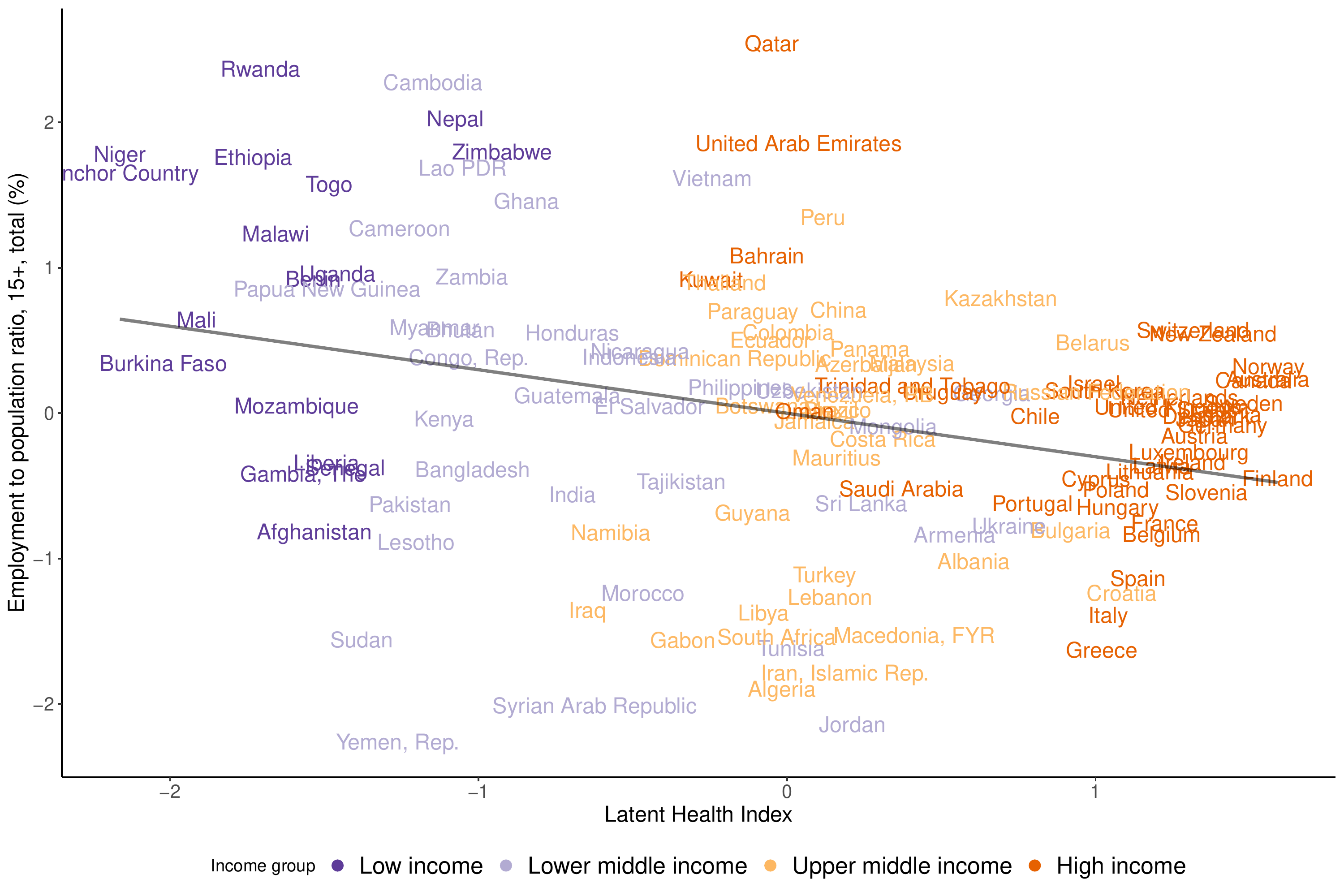} 
\label{fig:employ}
\end{figure}
\vspace{-1.5em}

For example, Figure \ref{fig:employ} shows that there is a weak negative relationship between the metric `\texttt{employment to population ratio in a country (for population aged 15 and over)}' and a country's latent health (95\% credible interval for $a_2$ is $(-0.49, -0.07)$). As we can see from the figure, countries with a high proportion of employment are in the low-income group, and the proportion decreases with the increasing level of income. This goes against the naive belief that a high employment rate reflects a country's `good' health. There are several possible explanations for this result. One, the population in high-income countries have higher life expectancy, and therefore, there is a higher proportion of retirees who are not employed. Two, a weak or absent social safety net in the low-income countries may require more people to work past retirement age.

This result also demonstrates that our model-based approach does not require \textit{a priori} input on which metrics reflect `good' health. We display similar results for two other metrics in Appendix \ref{appendix:appC}.

\subsubsection*{Nations' latent health, $H$}
Table \ref{tab:param1} shows the four highest ranked $H_i$ from model A, corresponding to Germany ($i$ = 28), Switzerland ($i$ = 20), USA ($i$ = 118) and Canada ($i$ = 19). Based on the 95\% credible intervals, Figure \ref{fig:HrankA} suggests that there are no stark differences from country to country. Nevertheless, we can consider potential groupings by examining the posterior probability of a positive difference in health between countries. For the four top-ranked countries, the posterior probability for Germany to be in better health than Switzerland is 0.95, whereas for Switzerland to be better than USA is 0.81. However, for USA and Canada, it is negligible with posterior probability equal to 0.53, suggesting that these two countries may be grouped together. Similar calculations of posterior probabilities can also be easily obtained for other countries.     

\subsubsection*{Inverse decay parameter for spatial correlation function, $\phi$}
\begin{figure}[H]
\caption{Spatial correlation function $\rho = exp(-\frac{d}{\phi})$ using posterior of $\phi$}
\includegraphics[height=0.3\textheight, width=0.85\textwidth]{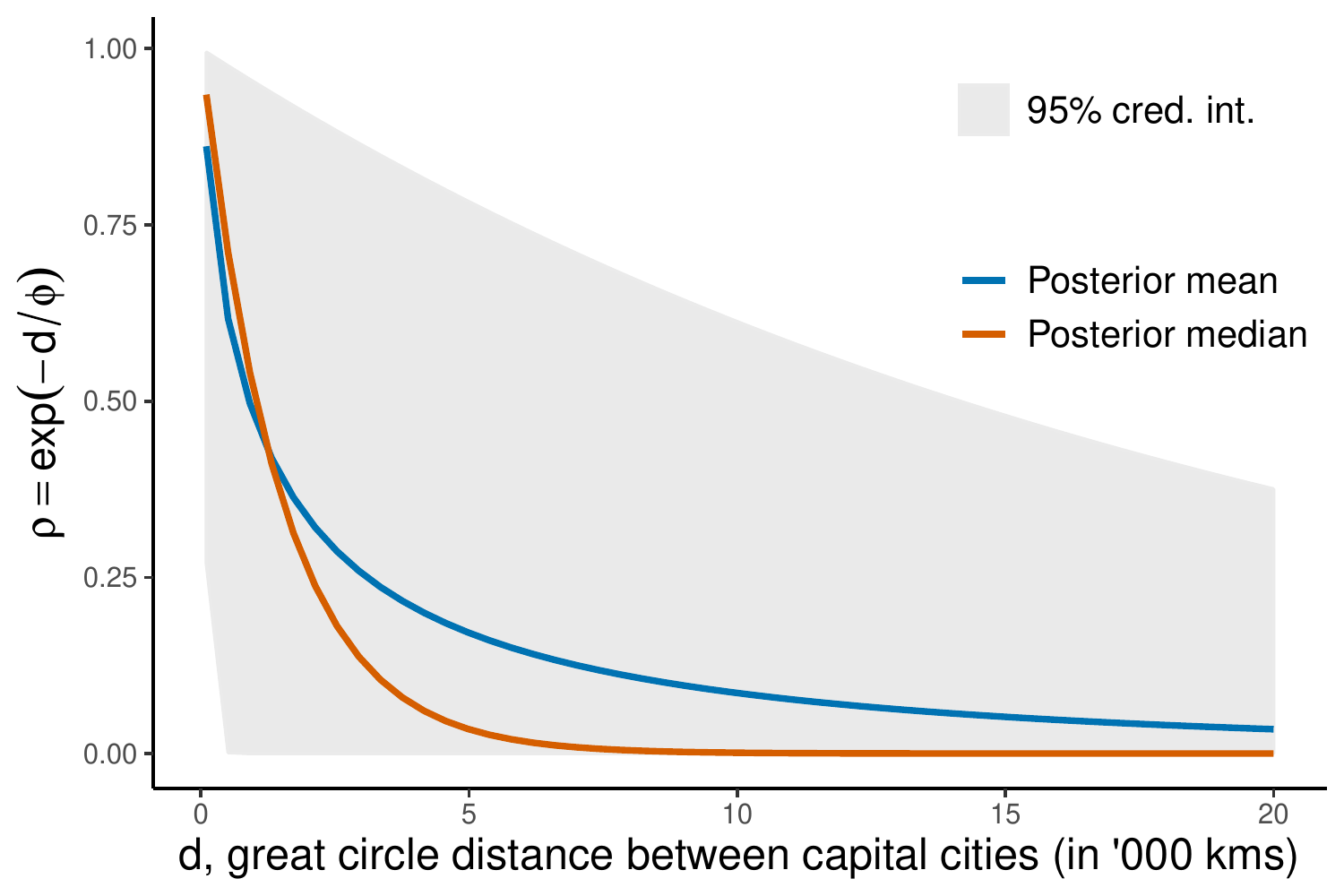}
\label{fig:rho_covar}
\end{figure}

Figure \ref{fig:rho_covar} shows the evaluated mean and median of the spatial correlation function $\rho$ calculated using posterior samples of $\phi$'s plotted against the great circle distance (GCD) between the capital cities.

\vspace{-1.5em}

As the inverse decay parameter $\phi$ has a right-skewed posterior distribution (trace plot in Table \ref{tab:param1}), we prefer the interpretation of $\rho$ using the posterior median, which suggests that the spatial correlation decreases quite sharply to 0 for distances up to 7,000km (approximately the GCD between capitals of USA and Estonia). Meanwhile, the upper bound of the 95\% credible interval of $\rho$ for two capitals that are the furthest apart (Spain and New Zealand) is 0.38. 

\begin{figure}[H]
	\caption{Posterior probability of spatial correlation function $\rho$ greater than selected threshold}
	\includegraphics[height=0.3\textheight, width=0.85\textwidth]{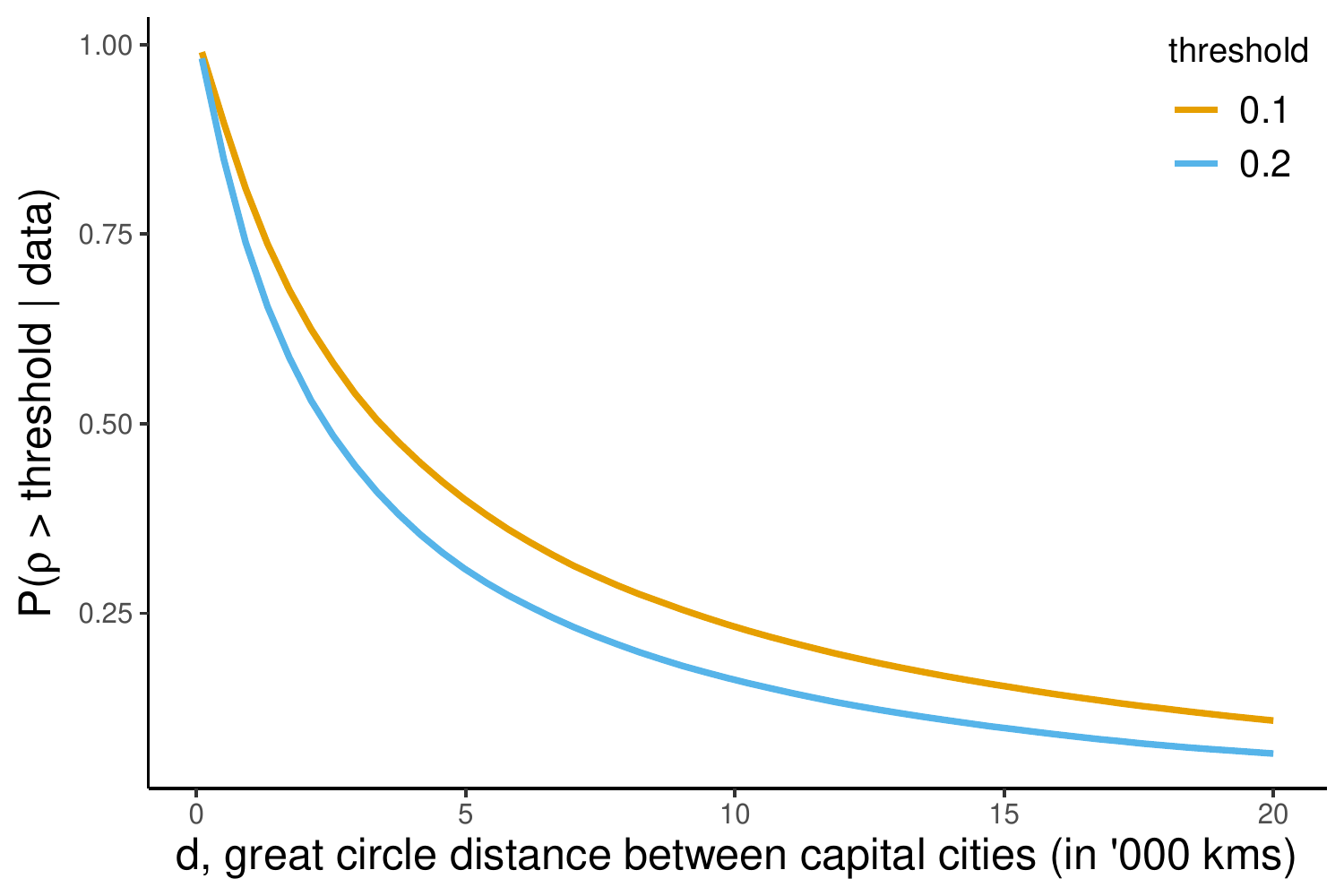}
	\label{fig:rho_thres}
\end{figure}

We also determined the posterior probability of $\rho$ being greater than the selected threshold of 0.1 or 0.2, presented in Figure \ref{fig:rho_thres}.

\vspace{-0.7em}
\vspace{-0.7em}

The posterior probability of $\rho > \text{threshold}$ in Figure \ref{fig:rho_thres} shows that for capital cities that are apart at approximately 5,000km in GCD (e.g. Vietnam and Papua New Guinea), the probability of a spatial correlation greater than 0.1 (0.2) is about 0.4 (0.3).

\subsubsection*{Non-causal policy treatment effect, $\beta_1$}
The regression coefficient $\beta_1$ in model A corresponds to the effect of MML days on the country's latent health as in a non-causal regression, while holding other covariates constant. The effect appears to be minor, with a 95\% credible interval that includes zero: see Table \ref{tab:param1}. In model B, we formally examine the effect of MML days as a  `policy treatment' using causal modeling.

\subsection{Model B: Spatial LHFI + causal modeling}\label{ssec:modelB}
To help answer policy prescription questions, we incorporate a formal causal modeling structure into our spatial LHFI modeling framework (model A). This is accomplished by including the treatment variable $T$ and a function of its propensity scores $z(\cdot)$, which are based on $K=4$ covariates as set out in Section \ref{sec:BPSA}.

\subsection*{Model B: Priors}
The priors for the parameters $\bm{\xi}$ and $\bm{\gamma}$ were diffuse normal distributions (mean 0 and variance 100). The same prior distributions from model A (Section \ref{ssec:modelA}) were also used for all other parameters here.

\subsubsection{Model B: Ranking of countries according to latent health, H}
\phantomsection{\hspace{0.5em}}

\begin{figure}[H]
\vspace{-1.5em}
	\caption{Model B: Latent health for 125 countries in 2015 grouped by income group}
	\includegraphics[width=\textwidth, height=0.7\textheight]{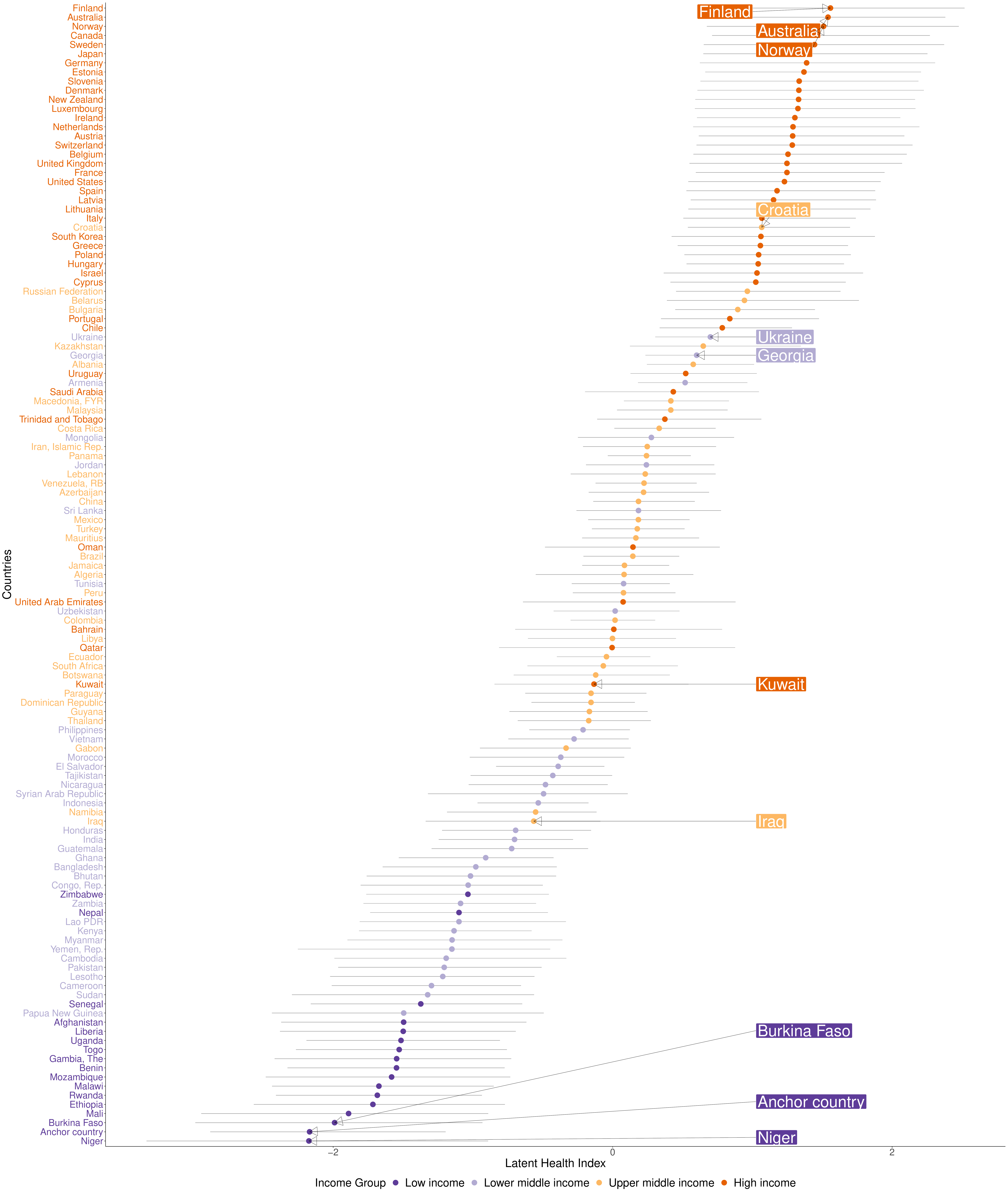} \label{fig:HrankB}
\end{figure}

In comparison to the latent health ranking from model A, Figure \ref{fig:HrankB} shows a higher uncertainty (wider gray bands) around the posterior median of health when including the PS inference (causal modeling) in the spatial LHFI model (model A). However, the overall ranking is fairly similar to the results from model A, with high income group countries ranked at the top, and low income group countries ranked at the bottom, but some exceptions in between. Two exceptions are Ukraine and Georgia, which are both classified as lower middle income countries by the United Nations, but they are ranked among the high and upper middle income countries in our rankings from both models A and B. In contrast, Kuwait, a high income country, and Iraq, an upper middle income country, are both ranked at the bottom half along with other lower middle and low income countries. These results again suggest that a country's health is not solely reflected by its income or wealth, aligning with the model A results of the metrics' loadings.

The posterior median and corresponding credible interval for the highest-ranked, lowest-ranked, and the anchor country are shown in Table \ref{tab:coefB}. Note that the countries ranked at the top (Finland) and bottom (Niger) are now different from model A (Germany and Burkina Faso, respectively).

\subsubsection*{Model B: Results}\label{ssec:coefB}
Posterior summaries for selected parameters from model B are shown in Table \ref{tab:coefB}. The other parameters are tabulated in Appendix \ref{appendix:appB}.

\FloatBarrier
\begin{table}[!htbp]
\caption{Posterior summaries for selected model B parameters}
\begin{tabular}{cccccc}
\toprule
Parameter & Markov Chain & 2.5\% & Median & 97.5\% & P($\beta_1 > 0$) \\
\midrule 
$H_{36\phantom{,} \text{(Finland)}}$ & \includegraphics[width=0.1\textwidth, height=0.03\textheight]{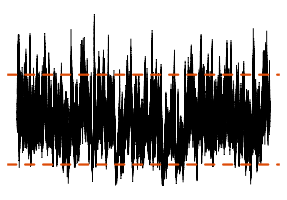} & \hlt{0.71} & 1.56 & \hlt{2.52} & - \\
$H_{84\phantom{,} \text{(Niger)}}$ & \includegraphics[width=0.1\textwidth, height=0.03\textheight]{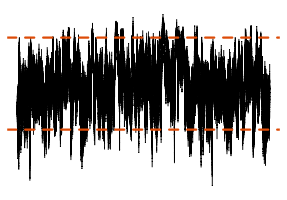} & \hlt{-3.34} & -2.17 & \hlt{-0.89} & -\\
$H_{anc}$ & \includegraphics[width=0.1\textwidth, height=0.03\textheight]{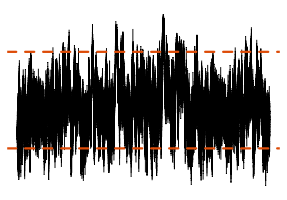} & \hlt{-2.88} & -2.17 & \hlt{-1.19} & -\\
$\beta_1$  & \includegraphics[width=0.1\textwidth, height=0.03\textheight]{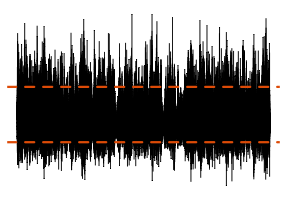}& \hlt{-0.14} & 0.11 & \hlt{0.48} &  0.82 \\
$\xi_1$ & \includegraphics[width=0.1\textwidth, height=0.02\textheight]{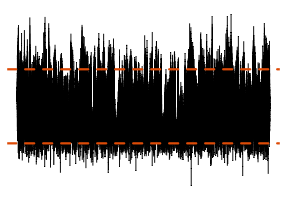} & \hlt{ 0.16} & 0.72                                                                                                                         &  \hlt{1.43} & - \\
$\xi_2$ &\includegraphics[width=0.1\textwidth, height=0.02\textheight]{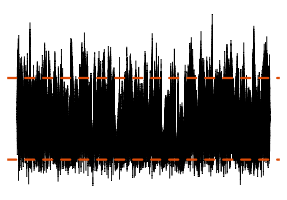} & \hlt{0.29} & 1.15 & \hlt{2.39} & - \\ 
\bottomrule
\end{tabular}
\label{tab:coefB}
\end{table}
\FloatBarrier

\subsubsection*{Policy treatment causal effect, $\beta_1$} 
\begin{figure}[htb!]
\caption{Trace plot of policy treatment effect}
\includegraphics[height=0.3\textheight, width=0.9\textwidth]{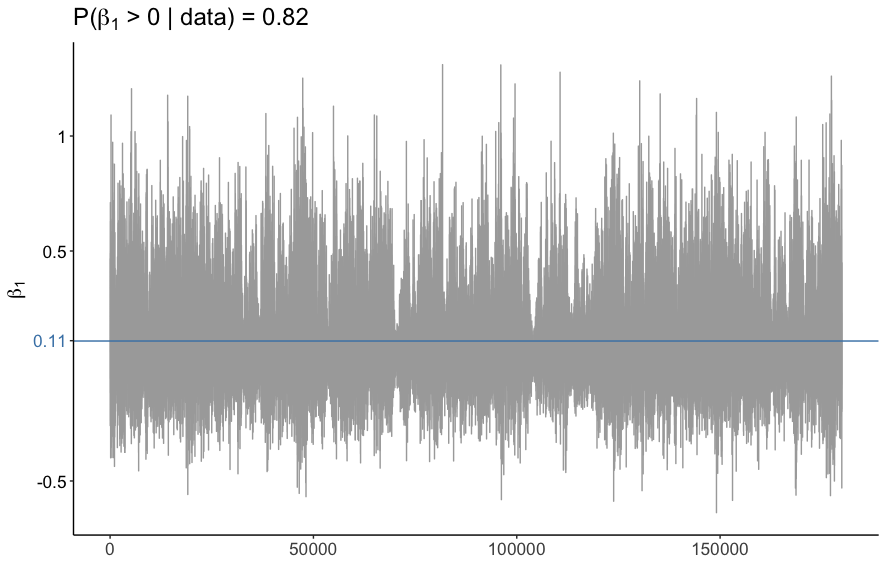}
\label{fig:beta1}
\end{figure}

Figure \ref{fig:beta1} shows that the posterior median of $\beta_1$ is 0.11, while the posterior probability of a positive policy treatment effect is 0.82. In addition, the posterior probability that the policy treatment effect is greater than the model A posterior median (0.03) is 0.74. Our findings for 2015 (and 2010--2014, not shown)\footnote{We obtain consistent findings about MML's effect on health based on model B for each of 2010--2014 with a varying number of countries, $N = \{104, 113, 100, 116, 108\}$. The different $N$'s are due to missingness in either $\boldsymbol{Y}$, $\boldsymbol{T}$, or $\boldsymbol{X}$. The posterior median policy treatment effect was $\{0.16, 0.11, 0.20, 0.11, 0.10\}$ with the posterior probability of a positive treatment effect being $\{0.90, 0.90, 0.92, 0.81, 0.85 \}$, respectively.} suggest a non-trivial positive treatment effect on a country's latent health. This modeled effect may be somewhat cruder than reality due to the dichotomization of the MML variable at the median, so that `treatment' refers to a federal policy being above the median number of MML days among all countries. In Section \ref{sec:future}, we discuss the potential to model the treatment variable as a non-dichotomized ordinal variable.

\vspace{-0.8em}
\vspace{-0.5em}

\subsubsection*{Regression coefficient $\bm{\xi}$ in the subclass indicator function $g(\cdot)$}
In model B, the propensity score is a function of four covariates ($X$). On the $H$-level, for each MCMC iteration, the propensity scores are grouped into tertile subclasses through the indicator function $g(\cdot)$ (see Section \ref{sec:BPSA}). The $\bm{\xi}$'s are the associated regression coefficients for the categorical variable with 3 levels. Note that the posterior medians for $\xi_1$ and $\xi_2$ shown in Table \ref{tab:coefB} are monotonically increasing. This suggests that countries with a higher propensity to receive the `policy treatment' tend to exhibit better health. 
 
\newpage
\subsubsection*{Indicator function $g(z{[}X,\bm{\gamma}{]})$ for subclass membership of propensity score} \label{ssec:PS}
\hfill 

\vspace{-1.7em}
\begin{figure}[H]
		\caption{$X$ vs. approximate subclass, for countries with below median MML days (light blue) and above median MML days (dark blue), where an approximate subclass is the tertile group in which the country exhibits the highest posterior probability of membership}
	\includegraphics[width=0.9\textwidth, height=0.38\textheight]{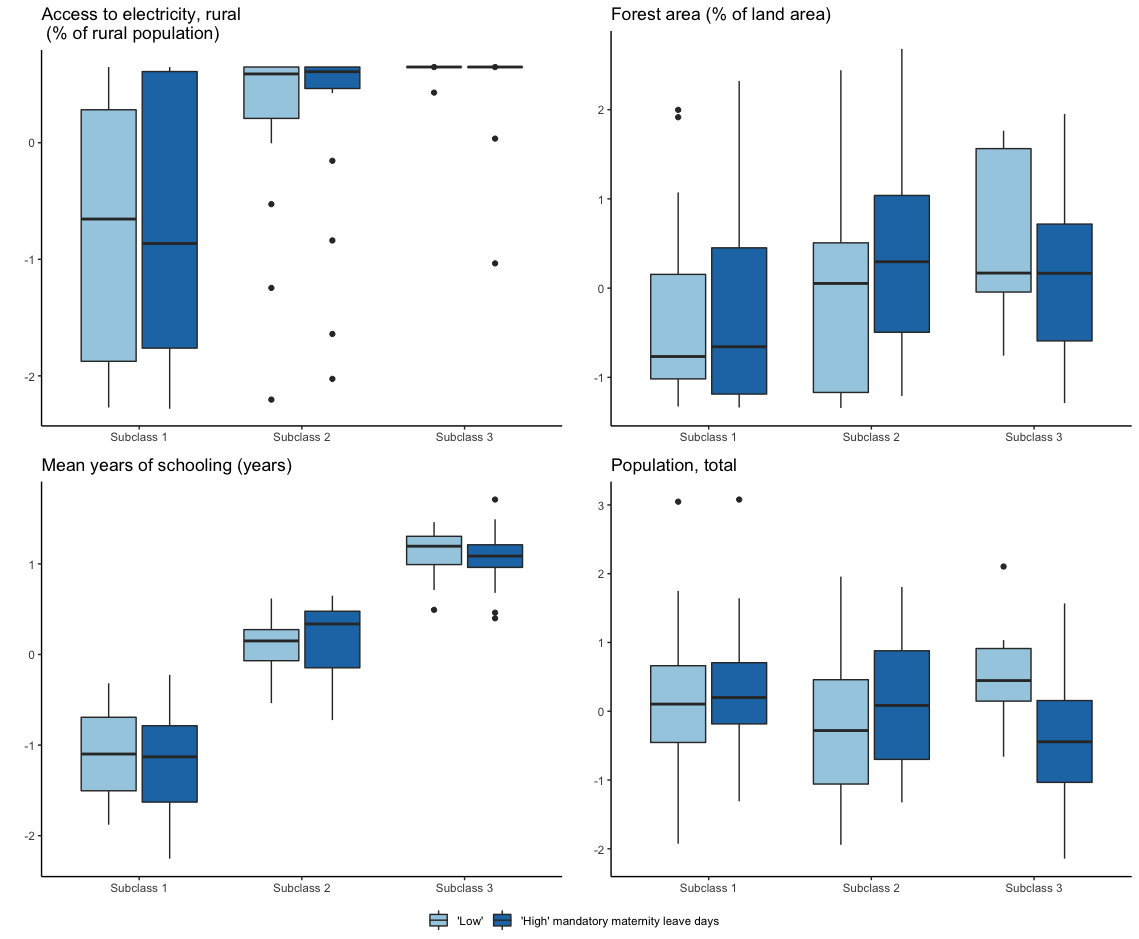}\label{fig:PSass} 
\end{figure}
\vspace{-1.5em}

In Rubin's approach of causal inference, we seek to achieve balance between the treatment and control groups in our study with respect to the pre-treatment covariates [\Mycite{imbens2015}]. To assess this balance, we consider Figure \ref{fig:PSass}, which shows the pre-treatment covariates in our case, namely the national level structural variables, plotted against the three approximate subclasses; by `approximate', we refer to a naive Bayes classification under which each country's post-burn-in MCMC samples of $g$ is tabulated, and the value of $g$ with the highest posterior frequency is regarded as the country's approximate subclass. 

Generally, the treated countries (above median MML days, in dark blue) are similar in group size compared to the controls (below median MML days, in light blue), suggesting covariate balance within the propensity subclasses; the only exception of covariate imbalance appears in the highest PS tertile for the \texttt{population, total} covariate. Thus, we have adequate overall covariate balance within each subclass, which, along with the assumption that conditioning on the existing infrastructure variables captures all key drivers of a country's health (so that the treatment assignment of each country to its MML days is strongly ignorable, see Section \ref{sec:BPSA}), allow the estimation of average policy treatment effect between the treated and control countries [\Mycite{rosenbaum1984}; \Mycite{imbens2015}]. 

\section{Discussion}\label{sec:discussion}
\subsection{Identifiability}\label{ssec:ident}
Recall that at the metric-level (Y-level), $a_j$ is the population-level loading of any country's health on its $j$th metric. However, in addition to the discussions by \Mycite{chiu2011a} and \Mycite{martin2002}, currently the multivariate metric vector $\bm{y_i}$ from a given year is modeled separately from other years, so that health $H_i$ being a random effect leads to an unidentifiable $\bm{a}$ unless constraints are imposed. 

One possible constraint is to prespecify the mean and variance for the health of an anchor country (e.g. $H_{anc} \sim N(-2, 0.1)$). To decide on the anchor country and its mean, we conducted a pilot run of the base LHFI model in Section \ref{sec:LHFI} but without any anchor, then selected a low-income group country on the extreme end of the $H$-scale as the anchor in all subsequent formal models.

As the ranking of the countries is relative, setting the constraint restricts the anchor country's health in the negative space and imposes this fixed scale on all other countries. This constraint solves the parameter identifiability issue along with aiding the interpretation of $H_i$, as it encodes in the model that a higher value of $H_i$ should be interpreted as a higher level of health, not lower. 

We have explored alternative constraints, including a `hard anchor' (fixing $H_{anc}$ to a constant), a `truncated anchor' (using a truncated multivariate normal distribution on the $H$-level), and transposing the $H$-scale manually post-MCMC sampling. Based on very preliminary results, we believe the `truncated anchor' to be the most desirable as it is flexible in restricting the anchor country to be in the negative space (or positive space, if desired) without specifying a particular mean and variance. While this constraint leads to additional computational burden, future iterations of this work will further explore this approach. 

\subsection{Spatial distances}\label{ssec:spatial}
In models A and B (Sections \ref{ssec:modelA} and \ref{ssec:modelB}), spatial correlation between countries was included to explicitly account for their respective geographical locations. It is modeled through the simplest Mat\'ern covariance function in the form of an exponential decay over great circle distances between capital cities of countries. Note that due to the earth's spherical nature, Euclidean distances may be inappropriate on a global scale [\Mycite{banerjee2005}]. Alternatively, we may explore other choices of the covariance function when we extend models A and B to more complex forms that formally incorporate dependencies jointly across time (2010--2015) and space. \Mycite{gleditsch2001} have also defined a minimum distance between countries based on country borders up to a certain distance. In our future work we may explore this distance measure. 

\subsection{Data transformation and missing values}\label{ssec:missing}
Higher unemployment rate is generally regarded as bad for societal health [\Mycite{wulfgramm2014}; \Mycite{helliwell2014}]. For this reason, the metric for the unemployment rate had been linearly transformed prior to modeling so that higher values reflect better societal health. The same transformation was also applied to the metric for infant mortality rate. In practice, the modeler needs not carry out this transformation, as the fitted model can be used to distinguish the strength and direction of the relationship between latent health and metrics, as reflected by the signs of the metrics' loadings. For instance, the results from both of our models suggest that higher values of latent health are associated with lower values of the unemployment metric (equivalent to more unemployment in the country after the data transformation). In other words, it is inferred that countries with better latent health have more unemployment, conditioned on the set of metrics and covariates that are included in our model. Similar implications of this result were discussed in Section \ref{ssec:metric}. Some visualizations and summaries of the posterior distribution for the negative metric effect, $a_j$, are included in Appendix \ref{appendix:appC}. 

Note that the selection of variables to be included as metrics and covariates in our model is largely based on the availability of data. While this paper focuses on the development of methodology, when applying the methodology in practice, the covariates and metrics could be specified by the modeler more according to their domain knowledge and less to data availability. In either case, the issue of missing data may require special attention.

In this paper, we presented the results for the year 2015. In addition, for model B, we also briefly discuss the policy treatment effect of MML days for the years 2010--2014. Our data for MML was obtained from the World Bank, which only collects MML data every other year. Therefore, data for the years 2010, 2012, and 2014 are considered missing. For those years, only countries with the same values for the years before and after the missing year entered our model. To further reduce missingness of the data in each year, for some of the OECD countries, the OECD data were used when there are missing MML values in the World Bank data, although we note that the two organizations have slightly different definitions of maternity leave.

Even if data exist in published records, it is recognized that such data collected on the country-level by various world organizations may have been derived from different and unpublished imputation techniques. Of course, the quality of the data would depend on the actual imputation techniques employed. Moreover, one may not rule out the possibility that data or official statistics reported by certain countries may have been fabricated. Although these disadvantages could reduce the accountability of our modeling results, overcoming such data-related challenges is beyond the scope of our paper.

\section{Future work}\label{sec:future}
As mentioned in Section \ref{sec:BPSA}, to facilitate formal causal inference, we have incorporated the Bayesian extension of the so-called Rubin's approach of subclassification of propensity score [\Mycite{mccandless2009}; \Mycite{zigler2014}] in the LHFI framework. In particular, we have used the rather ad-hoc method of stratification on tertiles of the Bayesian inference of the propensity scores. Formal matching on the `treated' and `untreated' countries [\Mycite{gelman2007}] is a potential alternative approach yet to be explored. In addition, we intend to consider the alternative framework of Pearl's causal diagram approach, which could reveal if, by controlling for certain variables, we have unintentionally opened some `backdoor paths' in the causal diagram which would result in spurious correlation [\Mycite{pearl2009}]. In the literature, backdoor paths are any non-causal paths between the treatment and outcome variables in the causal diagram. As such, Pearl's approach might prompt us to control for a different set of variables and potentially lead to a different scientific conclusion. 

Instead of using the continuous MML variable in our work, we have dichotomized it to higher (1) or lower (0) than the median number of days, thus transforming it into a binary treatment variable, as appeared in \Mycite{williamson2012} and \Mycite{rubin2005}. We intend to explore employing a continuous treatment variable, set out in \Mycite{hirano2004}, to fully utilize the MML data and incorporate all the raw information while reducing the ambiguity due to an arbitrary quantile used to categorize treatments. Note that while MML days was chosen as the policy treatment variable in our paper, another policy variable of interest could have been selected based upon, say, the availability of data. 

Because the publicly available data that we have collated appear in the form of annual records throughout 2010--2015, it would be reasonable to formally model the temporal correlation as an extension to this paper. This would result in a spatio-temporal hierarchical causal model. We anticipate that careful consideration of separability or otherwise between space and time will be required.

Lastly, we would be interested in the formal inference that allows us to identify which metrics are crucial in reflecting the health of a country. Thus, ideally, we would model the metric effects as proportions that sum to 1, resulting in a type of variable/model selection framework. The implication of such a parameterization is the reduction in any modeler-induced selection bias due to choosing variables that are \textit{a priori} perceived as being important in reflecting a country's health.

\section*{Acknowledgments}
This research is supported by an IBISWorld philantrophic donation by Phil Ruthven to the Australian National University in the form of research funds awarded to GS Chiu. We thank Beatrix Jones, Bruce Chapman, Carolyn Huston, Corwin Zigler, Peter Mueller, Tim Higgins, and the attendees of Bayes on the Beach 2017 and Joint Statistical Meeting 2018 for stimulating discussions and constructive comments on the topic.

\appendix
\section{MCMC algorithm}\label{appendix:MCMC}
All model inference is done through MCMC. All parameters with conjugate priors are updated via Gibbs sampling unless stated otherwise: 

\begin{enumerate}
	\item Sample $H_{i|-i}$ from N(M, V) full conditional distribution where: 
	$$ \text{V} = \left( \boldsymbol{a}^T \Sigma_{Y}^{-1} \boldsymbol{a} + D^{-1} \right) $$
	$$ \text{M} = \text{V}^{-1}\left( \boldsymbol{a}^T\Sigma_{Y}\boldsymbol{y_i} + D^{-1} m_i \right)$$
	$$ D = \Sigma_{H[i,i]} - \Sigma_{H[i,-i]}^T \Sigma_{H[-i,-i]}^{-1} \Sigma_{H[-i,i]} $$ 
	$$ m_i = \mu_i + \Sigma_{H[i,-i]} \Sigma_{H[-i,-i]}^{-1} (x_i - \mu_{-i})$$
	$$ \mu_i = [\mathbf{T}\boldsymbol{\beta}]_i \text{ -- \textbf{for model A}}$$
	$$ \mu_{-i} = [\mathbf{T}\boldsymbol{\beta}]_{-i} \text{ -- \textbf{for model A}}$$
		$$ \mu_i = [\mathbf{T}\boldsymbol{\beta} + \boldsymbol{g(z(X,\bm{\gamma}))\bm{\xi}}]_i \text{ -- \textbf{for model B}}$$
	$$ \mu_{-i} = [\mathbf{T}\boldsymbol{\beta} + \boldsymbol{g(z(X,\bm{\gamma}))\bm{\xi}}]_{-i} \text{ -- \textbf{for model B}}$$
	$$\text{for } i = 1, \ldots, N; i \neq \text{anc} $$
	
	\item Sample $\boldsymbol{a}$ from its MVN($\bm{M}$, $\bm{V}$) full conditional distribution where:
	$$ {\bm{V}} = \left( \sum_{i = 1}^{N} H_i^2\Sigma_{Y}^{-1} + \Lambda_a\right)$$
	$$ \bm{M} =  \bm{V}^{-1}  \left( \Sigma_{Y}^{-1} \mathbf{Y}^T \bm{H} \right) $$
	
	\item Sample $\Sigma_Y$ from its Inv-Wishart($\nu_n$, $\bm{S_n}$) full conditional distribution where:
	$$\nu_n = \nu_0 + N $$
	$$ \bm{S_n} = (\mathbf{Y} - \bm{H}\boldsymbol{a}^T)^T(\mathbf{Y} - \bm{H}\boldsymbol{a}^T)+ \bm{S_0} $$
	
	\item Sample $\sigma^2_H$ from its Inv-Gamma($\alpha_n, \beta_n$) full conditional distribution where:
	$$\alpha_n = N/2 + \alpha_H$$
	$$\beta_n = \sum_{i=1}^{N} D_i^2 / 2 + \beta_H$$
	$$ D_i = H_i -  X_i \bm{\beta} \text{ - \textbf{for model A}} $$
	$$ D_i = H_i - T_i \bm{\beta} - \bm{\xi}^T \bm{g}(\bm{z}(X_i,\bm{\gamma}))  \text{ - \textbf{for model B}} $$
	
	\item \textbf{(For model A only)} Sample $\bm{\beta} = \{\beta_0, \ldots, \beta_{6}\}$ from its MVN($\bm{M}$, $\bm{V}$) full conditional distribution where:
	$$ {\bm{V}} = \bm{X}^T \Sigma_H \bm{X} + \Lambda_{\beta}^{-1} $$
	$$ {\bm{M}} = {\bm{V}^{-1}} \left( \bm{X}^T \Sigma_H \bm{D}\right) $$
	$$ \bm{D} = \bm{H} $$
	
	\item \textbf{(For model B only)} Sample $\bm{\beta} = \{\beta_0, \beta_{1}\}$ from its MVN($\bm{M}$, $\bm{V}$) full conditional distribution where:
	$$ {\bm{V}} = \bm{X}^T \Sigma_H\bm{X} + \Lambda_{\beta}^{-1} $$
	$$ {\bm{M}} = {\bm{V}^{-1}} \left( \bm{X}^T \Sigma_H \bm{D} \right) $$
	$$ \bm{D} = \bm{H} - \bm{\xi}^T \bm{g}(\bm{z}(X_i,\bm{\gamma}))  $$
	
	\item Sample $\phi$ from its full conditional distribution using the Metropolis-Hastings (M-H) algorithm where: \\
	(\textbf{For model A only}): \\
$$ P(\phi | \bm{\beta}, \sigma_H^2) = \prod_{i = 1}^{N} p(H_i | \bm{\beta}, \sigma_H^2) p(\phi) $$ \\ 
where $\bm{\beta}$ is a 6 x 1 vector for 4 covariates, 1 treatment variable (included as a covariate here) and the intercept. \\

(\textbf{For model B only}): \\
$$ P(\phi | \bm{\beta}, \bm{\xi}, \sigma_H^2) = \prod_{i = 1}^{N} p(H_i | \bm{\beta}, \bm{\xi}, \sigma_H^2) p(\phi) $$ \\ 
where $\bm{\beta}$ is a 2 x 1 vector for 1 treatment variable, and the intercept. \\
	
	\item \textbf{(For model B only)} Sample $\bm{\xi}$ from its MVN($\bm{M}$, $\bm{V}$) full conditional distribution where: 
	$$ {\bm{V}} = \left(\bm{g}(\bm{z}(X_i,\bm{\gamma}))^T\Sigma_H^{-1}\bm{g}(\bm{z}(X_i,\bm{\gamma})) + \Lambda_\xi\right)$$
	$$ {\bm{M}} = {\bm{V}}^{-1} \left( \bm{g}(\bm{z}(X_i,\bm{\gamma})^T \Sigma_H^{-1} \bm{D}\right) $$
	$$ \bm{D} = \bm{H} - \bm{T}\bm{\beta}$$
	$$ \Sigma_H = \sigma_H^2 \mathbf{R}(d,\phi) \text{ as shown in equations (\ref{eq:3.32} - \ref{eq:spa5})}$$
	
	\item \textbf{(For model B only)} Sample $\bm{\gamma}$  using M-H algorithm from: 
	\begin{align*}
		p(\bm{\gamma} | T_i, X_i) &= \prod_{i = 1}^{N} p(T_i | X_i, \bm{\gamma})p(\bm{\gamma}) \\ 
	& = \prod_{i = 1}^{N} \frac{exp(T_i(X_i\bm{\gamma}))}{1 + exp(X_i\bm{\gamma})} p(\bm{\gamma})
	\end{align*}
\end{enumerate}
 
 \newpage
\section{Results for models A and B}
\subsection{Posterior summaries for model A}\label{appendix:appB} \phantomsection{\hspace{1em}}

\begin{savenotes}
\begin{table}[htb!]
	{\scriptsize
	\label{tab:paramA}
	\begin{tabular}{rrrr}
		\toprule
		$a_j$ & 2.5\% & 50\% & 97.5\% \\ 
		\midrule
		$a_1$ & 0.90 & 1.10 & 1.44 \\ 
		$a_2$ & -0.49 & -0.27 & -0.07 \\ 
		$a_3$ & 0.77 & 0.97 & 1.29 \\ 
		$a_4$ & 0.79 & 1.00 & 1.32 \\ 
		$a_5$ & 0.71 & 0.92 & 1.23 \\ 
		$a_6$ & 0.77 & 0.97 & 1.29 \\ 
		$a_7$ & -0.19 & 0.01 & 0.22 \\ 
		$a_8$ & 0.87 & 1.07 & 1.40 \\ 
		$a_9$ & 0.65 & 0.86 & 1.15 \\ 
		$a_{10}$ & 0.55 & 0.75 & 1.03 \\ 
		$a_{11}$ & -0.74 & -0.49 & -0.29 \\ 
		$a_{12}$ & -0.05 & 0.15 & 0.36 \\ 
		$a_{13}$ & -0.47 & -0.24 & -0.05 \\ 
		$a_{14}$ & 0.32 & 0.52 & 0.77 \\ 
		$a_{15}$ & 0.52 & 0.71 & 0.99 \\ 
		\bottomrule
	\end{tabular} 
	\begin{tabular}{rrrr}
	\hline
	$\beta_k$& 2.5\% & 50\% & 97.5\% \\ 
	\hline
	$\beta_{0}$ & -0.10 & -0.01 & 0.08 \\ 
	$\beta_{1}$ & -0.04 & 0.03 & 0.10 \\ 
	$\beta_{2}$ & 0.09 & 0.16 & 0.23 \\ 
	$\beta_{3}$ & -0.03 & 0.01 & 0.04 \\ 
	$\beta_{4}$ & 0.56 & 0.75 & 0.94 \\ 
	$\beta_{5}$ & -0.02 & 0.01 & 0.05 \\
  \hline
  $\sigma_H^2$ & 0.0043 & 0.013 &  0.028  \\
  $\phi$ & 0.08 & 1.48 & 20.41 \\
	\midrule
    $\Sigma_{y\{2,13\}}$\footnote{Only the top five in magnitude in terms of the posterior median are presented.} & 0.51 & 0.69 &  0.92  \\
     $\Sigma_{y\{11,14\}}$ & 0.27 & 0.41 &  0.59  \\
      $\Sigma_{y\{11,12\}}$ & 0.17 & 0.33 & 0.51   \\
       $\Sigma_{y\{9,11\}}$ &  0.20 & 0.31 & 0.44 \\
        $\Sigma_{y\{9,14\}}$ & 0.19 & 0.30 & 0.43  \\
		\bottomrule
\end{tabular}
}
\vspace{1em}

\centering
\begin{tabular}{rl}
  \toprule
$j$ & Metrics, Y\\ 
  \midrule
1 & Education index \\ 
  2 & Employment to popn. ratio, 15+, total (\%) \\ 
  3 & GNI per capita (2011 PPP\$) \\ 
  4 & Internet users (\% of popn.) \\ 
  5 & Life expectancy at birth, total (years) \\ 
  6 & Mortality rate, infant (per 1,000 live births) \\ 
  7 & Popn. density \\ 
  8 & {\scriptsize Popn. with at least some secondary education (\% ages 25 and older)} \\ 
  9 & Popn., ages 65 and older (\% of total) \\ 
  10 & Popn., urban (\% of total) \\ 
  11 & {\scriptsize Renewable energy consumption (\% of total final energy consumption)} \\ 
  12 & {\scriptsize Proportion of seats held by women in national parliaments (\%)} \\ 
  13 & Unemployment, total (\% of total labor force) \\ 
  14 & POLITY index \\ 
  15 & Corruption Perception Index \\ 
   \bottomrule
\end{tabular}
\vspace{1em}

{\small
\centering
\begin{tabular}{rl}
  \hline
$k$ & Covariates, X \\ 
  \hline  
1 & Mandatory maternity leave (dichotomized) \\ 
  2 & Access to electricity, rural (\% of rural population) \\ 
  3 & Forest area (\% of land area) \\ 
  4 & Mean years of schooling (years) \\ 
  5 & Population, total \\ 
   \hline
\end{tabular}
}
\end{table}
\end{savenotes}

	\begin{sidewaystable}	
	  \begin{tabular}[htp!]{cccc}
		\hline
		$H_i$\footnote[2]{See next page for indexing of countries} & 2.5\% & 50\% & 97.5\% \\  \hline
		$H_{1}$ & -1.71 & -1.36 & -1.02 \\ 
		$H_{2}$ & 0.28 & 0.42 & 0.62 \\ 
		$H_{3}$ & 0.24 & 0.45 & 0.66 \\ 
		$H_{4}$ & 0.46 & 0.66 & 0.88 \\ 
		$H_{5}$ & 0.83 & 1.11 & 1.43 \\ 
		$H_{6}$ & 0.75 & 1.00 & 1.26 \\ 
		$H_{7}$ & 0.36 & 0.53 & 0.77 \\ 
		$H_{anc}$ & -2.18 & -1.76 & -1.35 \\ 
		$H_{9}$ & 0.67 & 0.90 & 1.14 \\ 
		$H_{10}$ & -1.73 & -1.38 & -1.02 \\ 
		$H_{11}$ & -2.53 & -2.04 & -1.55 \\ 
		$H_{12}$ & -1.10 & -0.84 & -0.59 \\ 
		$H_{13}$ & 0.61 & 0.84 & 1.08 \\ 
		$H_{14}$ & 0.05 & 0.20 & 0.39 \\ 
		$H_{15}$ & 0.67 & 0.92 & 1.19 \\ 
		$H_{16}$ & -0.22 & -0.05 & 0.13 \\ 
		$H_{17}$ & -1.60 & -1.24 & -0.91 \\ 
		$H_{18}$ & -0.21 & -0.04 & 0.18 \\ 
		$H_{19}$ & 0.86 & 1.16 & 1.47 \\ 
		$H_{20}$ & 0.93 & 1.26 & 1.58 \\ 
		$H_{21}$ & 0.36 & 0.53 & 0.75 \\ 
		$H_{22}$ & -0.12 & 0.07 & 0.27 \\ 
		$H_{23}$ & -1.12 & -0.87 & -0.65 \\ 
		$H_{24}$ & -1.05 & -0.80 & -0.59 \\ 
		$H_{25}$ & -0.21 & -0.05 & 0.10 \\ 
		$H_{26}$ & -0.02 & 0.12 & 0.27 \\ 
		$H_{27}$ & 0.55 & 0.77 & 1.01 \\ 
		$H_{28}$ & 1.04 & 1.40 & 1.77 \\ 
		$H_{29}$ & 0.79 & 1.05 & 1.34 \\ 
		$H_{30}$ & -0.27 & -0.12 & 0.05 \\ 
		$H_{31}$ & -0.26 & -0.04 & 0.11 \\ 
		\hline
	\end{tabular}
	\begin{tabular}{cccc}
		\hline
		& 2.5\% & 50\% & 97.5\%  \\\hline
		$H_{32}$ & -0.16 & -0.00 & 0.15 \\ 
		$H_{33}$ & 0.36 & 0.52 & 0.76 \\ 
		$H_{34}$ & 0.81 & 1.09 & 1.38 \\ 
		$H_{35}$ & -2.07 & -1.67 & -1.26 \\ 
		$H_{36}$ & 0.78 & 1.05 & 1.35 \\ 
		$H_{37}$ & 0.64 & 0.87 & 1.11 \\ 
		$H_{38}$ & -0.44 & -0.25 & -0.09 \\ 
		$H_{39}$ & 0.83 & 1.12 & 1.43 \\ 
		$H_{40}$ & 0.66 & 0.92 & 1.19 \\ 
		$H_{41}$ & -0.61 & -0.43 & -0.26 \\ 
		$H_{42}$ & -1.84 & -1.47 & -1.11 \\ 
		$H_{43}$ & 0.43 & 0.60 & 0.81 \\ 
		$H_{44}$ & -0.82 & -0.60 & -0.42 \\ 
		$H_{45}$ & -0.24 & -0.08 & 0.08 \\ 
		$H_{46}$ & -0.87 & -0.64 & -0.45 \\ 
		$H_{47}$ & 0.59 & 0.79 & 1.03 \\ 
		$H_{48}$ & 0.67 & 0.90 & 1.14 \\ 
		$H_{49}$ & -0.39 & -0.18 & -0.01 \\ 
		$H_{50}$ & -0.71 & -0.50 & -0.31 \\ 
		$H_{51}$ & 0.78 & 1.04 & 1.34 \\ 
		$H_{52}$ & 0.17 & 0.33 & 0.52 \\ 
		$H_{53}$ & -0.69 & -0.47 & -0.30 \\ 
		$H_{54}$ & 0.75 & 1.03 & 1.31 \\ 
		$H_{55}$ & 0.43 & 0.60 & 0.83 \\ 
		$H_{56}$ & 0.06 & 0.21 & 0.37 \\ 
		$H_{57}$ & 0.20 & 0.37 & 0.54 \\ 
		$H_{58}$ & 0.77 & 1.05 & 1.35 \\ 
		$H_{59}$ & 0.58 & 0.82 & 1.10 \\ 
		$H_{60}$ & -1.08 & -0.85 & -0.62 \\ 
		$H_{61}$ & -1.38 & -1.08 & -0.79 \\ 
		$H_{62}$ & 0.72 & 0.98 & 1.26 \\ 
		\hline
	\end{tabular}
	\begin{tabular}{cccc}
		\hline
		& 2.5\% & 50\% & 97.5\%  \\\hline
		$H_{63}$ & -0.45 & -0.26 & -0.04 \\ 
		$H_{64}$ & -1.04 & -0.77 & -0.55 \\ 
		$H_{65}$ & -0.17 & -0.03 & 0.10 \\ 
		$H_{66}$ & -1.68 & -1.34 & -1.01 \\ 
		$H_{67}$ & -0.38 & -0.20 & -0.05 \\ 
		$H_{68}$ & 0.32 & 0.53 & 0.75 \\ 
		$H_{69}$ & -1.25 & -0.96 & -0.71 \\ 
		$H_{70}$ & 0.80 & 1.09 & 1.39 \\ 
		$H_{71}$ & 0.69 & 0.93 & 1.19 \\ 
		$H_{72}$ & 0.78 & 1.07 & 1.36 \\ 
		$H_{73}$ & -0.93 & -0.66 & -0.43 \\ 
		$H_{74}$ & -0.08 & 0.08 & 0.25 \\ 
		$H_{75}$ & 0.12 & 0.32 & 0.47 \\ 
		$H_{76}$ & -2.22 & -1.77 & -1.35 \\ 
		$H_{77}$ & -1.28 & -0.99 & -0.73 \\ 
		$H_{78}$ & 0.09 & 0.29 & 0.53 \\ 
		$H_{79}$ & -1.94 & -1.55 & -1.18 \\ 
		$H_{80}$ & 0.06 & 0.22 & 0.42 \\ 
		$H_{81}$ & -1.69 & -1.36 & -1.04 \\ 
		$H_{82}$ & 0.27 & 0.44 & 0.64 \\ 
		$H_{83}$ & -0.89 & -0.67 & -0.44 \\ 
		$H_{84}$ & -2.50 & -1.99 & -1.52 \\ 
		$H_{85}$ & -0.85 & -0.64 & -0.43 \\ 
		$H_{86}$ & 0.75 & 1.00 & 1.27 \\ 
		$H_{87}$ & 0.79 & 1.07 & 1.37 \\ 
		$H_{88}$ & -1.15 & -0.89 & -0.64 \\ 
		$H_{89}$ & 0.78 & 1.06 & 1.36 \\ 
		$H_{90}$ & 0.02 & 0.20 & 0.37 \\ 
		$H_{91}$ & -1.16 & -0.88 & -0.64 \\ 
		$H_{92}$ & 0.13 & 0.28 & 0.44 \\ 
		$H_{93}$ & -0.07 & 0.09 & 0.25 \\ 
		\hline
	\end{tabular}
	\begin{tabular}{cccc}
		\hline
		& 2.5\% & 50\% & 97.5\%  \\\hline
		$H_{94}$ & 0.03 & 0.19 & 0.35 \\ 
		$H_{95}$ & -1.72 & -1.36 & -1.02 \\ 
		$H_{96}$ & 0.68 & 0.93 & 1.18 \\ 
		$H_{97}$ & 0.15 & 0.29 & 0.48 \\ 
		$H_{98}$ & -0.23 & -0.03 & 0.13 \\ 
		$H_{99}$ & 0.16 & 0.32 & 0.52 \\ 
		$H_{100}$ & 0.68 & 0.93 & 1.20 \\ 
		$H_{101}$ & -1.77 & -1.44 & -1.10 \\ 
		$H_{102}$ & 0.19 & 0.35 & 0.60 \\ 
		$H_{103}$ & -1.84 & -1.47 & -1.11 \\ 
		$H_{104}$ & -1.88 & -1.50 & -1.13 \\ 
		$H_{105}$ & -0.69 & -0.50 & -0.33 \\ 
		$H_{106}$ & 0.74 & 1.00 & 1.28 \\ 
		$H_{107}$ & 0.76 & 1.02 & 1.31 \\ 
		$H_{108}$ & -1.07 & -0.80 & -0.55 \\ 
		$H_{109}$ & -1.41 & -1.11 & -0.82 \\ 
		$H_{110}$ & -0.30 & -0.13 & 0.02 \\ 
		$H_{111}$ & 0.11 & 0.35 & 0.57 \\ 
		$H_{112}$ & 0.43 & 0.61 & 0.85 \\ 
		$H_{113}$ & -0.40 & -0.23 & -0.07 \\ 
		$H_{114}$ & -0.22 & -0.07 & 0.11 \\ 
		$H_{115}$ & -1.29 & -1.03 & -0.77 \\ 
		$H_{116}$ & 0.52 & 0.73 & 0.96 \\ 
		$H_{117}$ & -0.04 & 0.12 & 0.28 \\ 
		$H_{118}$ & 0.86 & 1.17 & 1.49 \\ 
		$H_{119}$ & 0.38 & 0.62 & 0.85 \\ 
		$H_{120}$ & 0.22 & 0.40 & 0.58 \\ 
		$H_{121}$ & -0.12 & 0.03 & 0.19 \\ 
		$H_{122}$ & -1.75 & -1.37 & -1.00 \\ 
		$H_{123}$ & 0.15 & 0.32 & 0.51 \\ 
		$H_{124}$ & -0.98 & -0.75 & -0.53 \\ 
		$H_{125}$ & -0.64 & -0.43 & -0.26 \\ 
		\hline
	\end{tabular}
\end{sidewaystable}

\begin{sidewaystable}
\begin{tabular}[htb!]{rl}
	 \hline
	$i$ & Country \\ 
	\hline
	1 & Afghanistan \\ 
	2 & Albania \\ 
	3 & United Arab Emirates \\ 
	4 & Armenia \\ 
	5 & Australia \\ 
	6 & Austria \\ 
	7 & Azerbaijan \\ 
	8 & Anchor country \\ 
	9 & Belgium \\ 
	10 & Benin \\ 
	11 & Burkina Faso \\ 
	12 & Bangladesh \\ 
	13 & Bulgaria \\ 
	14 & Bahrain \\ 
	15 & Belarus \\ 
	16 & Brazil \\ 
	17 & Bhutan \\ 
	18 & Botswana \\ 
	19 & Canada \\ 
	20 & Switzerland \\ 
	21 & Chile \\ 
	22 & China \\ 
	23 & Cameroon \\ 
	24 & Congo, Rep. \\ 
	25 & Colombia \\ 
	26 & Costa Rica \\ 
	27 & Cyprus \\ 
	28 & Germany \\ 
	29 & Denmark \\ 
	30 & Dominican Republic \\ 
	31 & Algeria \\
	\hline
\end{tabular}
\begin{tabular}{rl}
	\hline
	$i$ & Country \\ 
	\hline
	  32 & Ecuador \\ 
33 & Spain \\ 
34 & Estonia \\ 
35 & Ethiopia \\ 
36 & Finland \\ 
37 & France \\ 
38 & Gabon \\ 
39 & United Kingdom \\ 
40 & Georgia \\ 
41 & Ghana \\ 
42 & Gambia, The \\ 
43 & Greece \\ 
44 & Guatemala \\ 
45 & Guyana \\ 
46 & Honduras \\ 
47 & Croatia \\ 
48 & Hungary \\ 
49 & Indonesia \\ 
50 & India \\ 
51 & Ireland \\ 
52 & Iran, Islamic Rep. \\ 
53 & Iraq \\ 
54 & Israel \\ 
55 & Italy \\ 
56 & Jamaica \\ 
57 & Jordan \\ 
58 & Japan \\ 
59 & Kazakhstan \\ 
60 & Kenya \\ 
61 & Cambodia \\ 
62 & South Korea \\ 
	\hline
\end{tabular}
\begin{tabular}{rl}
	\hline
	$i$ & Country \\ 
	\hline
  63 & Kuwait \\ 
64 & Lao PDR \\ 
65 & Lebanon \\ 
66 & Liberia \\ 
67 & Libya \\ 
68 & Sri Lanka \\ 
69 & Lesotho \\ 
70 & Lithuania \\ 
71 & Luxembourg \\ 
72 & Latvia \\ 
73 & Morocco \\ 
74 & Mexico \\ 
75 & Macedonia, FYR \\ 
76 & Mali \\ 
77 & Myanmar \\ 
78 & Mongolia \\ 
79 & Mozambique \\ 
80 & Mauritius \\ 
81 & Malawi \\ 
82 & Malaysia \\ 
83 & Namibia \\ 
84 & Niger \\ 
85 & Nicaragua \\ 
86 & Netherlands \\ 
87 & Norway \\ 
88 & Nepal \\ 
89 & New Zealand \\ 
90 & Oman \\ 
91 & Pakistan \\ 
92 & Panama \\ 
93 & Peru \\ 
	\hline
\end{tabular}
\begin{tabular}{rl}
	\hline
	$i$ & Country \\ 
	\hline
	94 & Philippines \\ 
	95 & Papua New Guinea \\ 
	96 & Poland \\ 
	97 & Portugal \\ 
	98 & Paraguay \\ 
	99 & Qatar \\ 
	100 & Russian Federation \\ 
	101 & Rwanda \\ 
	102 & Saudi Arabia \\ 
	103 & Sudan \\ 
	104 & Senegal \\ 
	105 & El Salvador \\ 
	106 & Slovenia \\ 
	107 & Sweden \\ 
	108 & Syrian Arab Republic \\ 
	109 & Togo \\ 
	110 & Thailand \\ 
	111 & Tajikistan \\ 
	112 & Trinidad and Tobago \\ 
	113 & Tunisia \\ 
	114 & Turkey \\ 
	115 & Uganda \\ 
	116 & Ukraine \\ 
	117 & Uruguay \\ 
	118 & United States \\ 
	119 & Uzbekistan \\ 
	120 & Venezuela, RB \\ 
	121 & Vietnam \\ 
	122 & Yemen, Rep. \\ 
	123 & South Africa \\ 
	124 & Zambia \\ 
	125 & Zimbabwe \\ 
	\hline
\end{tabular}
\end{sidewaystable}

\clearpage

\subsection{Posterior summaries for model B}\label{ssec:appB2} \phantomsection{\hspace{1em}}

\begin{savenotes}
\begin{table}[hbt!]
	\label{tab:paramB}
	\centering
	\begin{tabular}{rrrr}
		\toprule
		$a_j$\footnote{Refer to appendix \ref{appendix:appB} for indexing of metrics.} & 2.5\% & 50\% & 97.5\% \\
		\midrule
$a_1$ & 0.61 & 0.90 & 1.74 \\ 
$a_2$ & -0.73 & -0.31 & -0.10 \\ 
$a_3$ & 0.55 & 0.84 & 1.62 \\ 
$a_4$ & 0.57 & 0.85 & 1.66 \\ 
$a_5$ & 0.55 & 0.83 & 1.65 \\ 
$a_6$ & 0.59 & 0.87 & 1.71 \\ 
$a_7$ & -0.27 & -0.03 & 0.18 \\ 
$a_8$ & 0.56 & 0.84 & 1.64 \\ 
$a_9$ & 0.51 & 0.77 & 1.61 \\ 
$a_{10}$ & 0.42 & 0.68 & 1.33 \\ 
$a_{11}$ & -0.83 & -0.39 & -0.16 \\ 
$a_{12}$ & -0.04 & 0.15 & 0.44 \\ 
$a_{13}$ & -0.62 & -0.26 & -0.06 \\ 
$a_{14}$ & 0.28 & 0.49 & 1.06 \\ 
$a_{15}$ & 0.43 & 0.65 & 1.25 \\ 
		\bottomrule
	\end{tabular}
   \vspace{1em}
	\begin{tabular}{rrrr}
		\toprule 
		& 2.5\% & 50\% & 97.5\% \\
		\midrule
    $\beta_0$ & -1.43 & -0.69 & 0.34 \\
    $\sigma_H^2$  & 0.15 & 0.55 & 1.66 \\
    $\phi$ & 0.07 & 1.29 & 15.29 \\  
		\midrule
		$\gamma_1$ & -0.68 & -0.14 & 0.42 \\
		$\gamma_2$ & -0.27 & 0.10 & 0.49 \\ 
		$\gamma_3$ & 0.20 & 0.76 & 1.34\\ 
		$\gamma_4$ & -0.41 & -0.02 & 0.37 \\ 
    \midrule
    $\Sigma_{y\{2,13\}}$\footnote{Only the top five in magnitude in terms of the posterior median are presented.} & 0.47 & 0.65 &  0.88  \\
     $\Sigma_{y\{11,14\}}$ & 0.27 & 0.42 &  0.60  \\
      $\Sigma_{y\{11,12\}}$ & 0.18 & 0.33 & 0.52   \\
       $\Sigma_{y\{9,11\}}$ &  0.18 & 0.31 & 0.45 \\
        $\Sigma_{y\{14,15\}}$ & 0.092 & 0.23 & 0.39  \\
		\bottomrule
	\end{tabular}
\end{table}
\end{savenotes}

  \begin{sidewaystable}
	  \begin{tabular}[H]{cccc}
		\toprule
		$H_i$ \footnote[5]{Refer to appendix \ref{appendix:appB} for indexing of countries} & 2.5\% & 50\% & 97.5\% \\  \midrule
$H_{1}$ & -2.37 & -1.50 & -0.62 \\ 
$H_{2}$ & 0.24 & 0.58 & 1.01 \\ 
$H_{3}$ & -0.64 & 0.07 & 0.88 \\ 
$H_{4}$ & 0.18 & 0.52 & 0.96 \\ 
$H_{5}$ & 0.67 & 1.54 & 2.38 \\ 
$H_{6}$ & 0.62 & 1.29 & 2.09 \\ 
$H_{7}$ & -0.17 & 0.22 & 0.69 \\ 
$H_{anc}$ & -2.88 & -2.17 & -1.19 \\ 
$H_{9}$ & 0.58 & 1.25 & 2.11 \\ 
$H_{10}$ & -2.33 & -1.55 & -0.77 \\ 
$H_{11}$ & -2.99 & -1.99 & -0.93 \\ 
$H_{12}$ & -1.65 & -0.98 & -0.40 \\ 
$H_{13}$ & 0.45 & 0.90 & 1.45 \\ 
$H_{14}$ & -0.70 & 0.01 & 0.78 \\ 
$H_{15}$ & 0.39 & 0.94 & 1.76 \\ 
$H_{16}$ & -0.21 & 0.14 & 0.48 \\ 
$H_{17}$ & -1.76 & -1.02 & -0.40 \\ 
$H_{18}$ & -0.71 & -0.12 & 0.41 \\ 
$H_{19}$ & 0.71 & 1.47 & 2.27 \\ 
$H_{20}$ & 0.60 & 1.29 & 2.15 \\ 
$H_{21}$ & 0.34 & 0.78 & 1.28 \\ 
$H_{22}$ & -0.14 & 0.18 & 0.59 \\ 
$H_{23}$ & -2.01 & -1.30 & -0.66 \\ 
$H_{24}$ & -1.80 & -1.03 & -0.50 \\ 
$H_{25}$ & -0.30 & 0.02 & 0.30 \\ 
$H_{26}$ & 0.01 & 0.33 & 0.74 \\ 
$H_{27}$ & 0.41 & 1.02 & 1.67 \\ 
$H_{28}$ & 0.62 & 1.39 & 2.31 \\ 
$H_{29}$ & 0.61 & 1.33 & 2.23 \\ 
$H_{30}$ & -0.58 & -0.15 & 0.16 \\ 
$H_{31}$ & -0.55 & 0.08 & 0.58 \\ 
		\bottomrule
	\end{tabular}
	\begin{tabular}{cccc}
		\toprule
		& 2.5\% & 50\% & 97.5\%  \\ \midrule
$H_{32}$ & -0.40 & -0.04 & 0.27 \\ 
$H_{33}$ & 0.53 & 1.18 & 1.88 \\ 
$H_{34}$ & 0.66 & 1.37 & 2.21 \\ 
$H_{35}$ & -2.57 & -1.72 & -0.77 \\ 
$H_{36}$ & 0.71 & 1.56 & 2.52 \\ 
$H_{37}$ & 0.59 & 1.25 & 1.95 \\ 
$H_{38}$ & -0.95 & -0.33 & 0.13 \\ 
$H_{39}$ & 0.55 & 1.25 & 2.07 \\ 
$H_{40}$ & 0.23 & 0.60 & 1.17 \\ 
$H_{41}$ & -1.53 & -0.91 & -0.42 \\ 
$H_{42}$ & -2.42 & -1.55 & -0.73 \\ 
$H_{43}$ & 0.46 & 1.06 & 1.68 \\ 
$H_{44}$ & -1.30 & -0.72 & -0.18 \\ 
$H_{45}$ & -0.74 & -0.17 & 0.25 \\ 
$H_{46}$ & -1.22 & -0.69 & -0.15 \\ 
$H_{47}$ & 0.54 & 1.07 & 1.70 \\ 
$H_{48}$ & 0.53 & 1.04 & 1.66 \\ 
$H_{49}$ & -0.97 & -0.53 & -0.17 \\ 
$H_{50}$ & -1.25 & -0.70 & -0.28 \\ 
$H_{51}$ & 0.60 & 1.30 & 2.06 \\ 
$H_{52}$ & -0.21 & 0.25 & 0.74 \\ 
$H_{53}$ & -1.34 & -0.56 & -0.09 \\ 
$H_{54}$ & 0.36 & 1.03 & 1.79 \\ 
$H_{55}$ & 0.51 & 1.07 & 1.74 \\ 
$H_{56}$ & -0.22 & 0.08 & 0.40 \\ 
$H_{57}$ & -0.19 & 0.24 & 0.73 \\ 
$H_{58}$ & 0.65 & 1.39 & 2.25 \\ 
$H_{59}$ & 0.12 & 0.65 & 1.35 \\ 
$H_{60}$ & -1.82 & -1.14 & -0.58 \\ 
$H_{61}$ & -1.99 & -1.19 & -0.33 \\ 
$H_{62}$ & 0.42 & 1.06 & 1.88 \\ 
		\bottomrule
	\end{tabular}
	\begin{tabular}{cccc}
			\toprule
			& 2.5\% & 50\% & 97.5\%  \\ \midrule
$H_{63}$ & -0.85 & -0.13 & 0.54 \\ 
$H_{64}$ & -1.81 & -1.10 & -0.33 \\ 
$H_{65}$ & -0.30 & 0.23 & 0.74 \\ 
$H_{66}$ & -2.38 & -1.50 & -0.69 \\ 
$H_{67}$ & -0.61 & -0.00 & 0.45 \\ 
$H_{68}$ & -0.26 & 0.18 & 0.78 \\ 
$H_{69}$ & -2.02 & -1.22 & -0.56 \\ 
$H_{70}$ & 0.54 & 1.12 & 1.85 \\ 
$H_{71}$ & 0.59 & 1.33 & 2.17 \\ 
$H_{72}$ & 0.56 & 1.15 & 1.88 \\ 
$H_{73}$ & -1.02 & -0.37 & 0.08 \\ 
$H_{74}$ & -0.18 & 0.18 & 0.55 \\ 
$H_{75}$ & 0.08 & 0.42 & 0.83 \\ 
$H_{76}$ & -2.94 & -1.89 & -0.89 \\ 
$H_{77}$ & -1.90 & -1.15 & -0.36 \\ 
$H_{78}$ & -0.25 & 0.28 & 0.87 \\ 
$H_{79}$ & -2.48 & -1.58 & -0.73 \\ 
$H_{80}$ & -0.22 & 0.17 & 0.62 \\ 
$H_{81}$ & -2.44 & -1.67 & -0.85 \\ 
$H_{82}$ & 0.03 & 0.42 & 0.82 \\ 
$H_{83}$ & -1.19 & -0.55 & -0.12 \\ 
$H_{84}$ & -3.34 & -2.17 & -0.89 \\ 
$H_{85}$ & -1.03 & -0.48 & -0.04 \\ 
$H_{86}$ & 0.58 & 1.29 & 2.19 \\ 
$H_{87}$ & 0.67 & 1.51 & 2.48 \\ 
$H_{88}$ & -1.74 & -1.10 & -0.46 \\ 
$H_{89}$ & 0.59 & 1.33 & 2.16 \\ 
$H_{90}$ & -0.48 & 0.15 & 0.77 \\ 
$H_{91}$ & -1.96 & -1.21 & -0.51 \\ 
$H_{92}$ & -0.04 & 0.24 & 0.56 \\ 
$H_{93}$ & -0.29 & 0.08 & 0.45 \\ 
			\bottomrule
		\end{tabular}
		\begin{tabular}{cccc}
			\toprule
			& 2.5\% & 50\% & 97.5\%  \\ \midrule
$H_{94}$ & -0.60 & -0.21 & 0.12 \\ 
$H_{95}$ & -2.44 & -1.50 & -0.49 \\ 
$H_{96}$ & 0.51 & 1.04 & 1.71 \\ 
$H_{97}$ & 0.35 & 0.84 & 1.48 \\ 
$H_{98}$ & -0.63 & -0.15 & 0.24 \\ 
$H_{99}$ & -0.81 & -0.00 & 0.88 \\ 
$H_{100}$ & 0.45 & 0.96 & 1.63 \\ 
$H_{101}$ & -2.41 & -1.68 & -0.94 \\ 
$H_{102}$ & -0.20 & 0.43 & 1.05 \\ 
$H_{103}$ & -2.30 & -1.32 & -0.56 \\ 
$H_{104}$ & -2.16 & -1.37 & -0.65 \\ 
$H_{105}$ & -0.83 & -0.39 & -0.06 \\ 
$H_{106}$ & 0.63 & 1.33 & 2.19 \\ 
$H_{107}$ & 0.65 & 1.44 & 2.37 \\ 
$H_{108}$ & -1.32 & -0.49 & 0.11 \\ 
$H_{109}$ & -2.27 & -1.53 & -0.76 \\ 
$H_{110}$ & -0.68 & -0.17 & 0.27 \\ 
$H_{111}$ & -1.02 & -0.43 & -0.00 \\ 
$H_{112}$ & -0.11 & 0.37 & 1.06 \\ 
$H_{113}$ & -0.29 & 0.08 & 0.41 \\ 
$H_{114}$ & -0.15 & 0.18 & 0.52 \\ 
$H_{115}$ & -2.19 & -1.51 & -0.81 \\ 
$H_{116}$ & 0.30 & 0.70 & 1.25 \\ 
$H_{117}$ & 0.13 & 0.52 & 1.03 \\ 
$H_{118}$ & 0.54 & 1.23 & 1.92 \\ 
$H_{119}$ & -0.42 & 0.02 & 0.48 \\ 
$H_{120}$ & -0.12 & 0.22 & 0.60 \\ 
$H_{121}$ & -0.75 & -0.28 & 0.12 \\ 
$H_{122}$ & -2.25 & -1.15 & -0.45 \\ 
$H_{123}$ & -0.61 & -0.07 & 0.47 \\ 
$H_{124}$ & -1.78 & -1.09 & -0.55 \\ 
$H_{125}$ & -1.76 & -1.04 & -0.46 \\ 
			\bottomrule
		\end{tabular}
  \end{sidewaystable}
\clearpage

\section{Some interesting results on parameters}\label{appendix:appC} 
\begin{figure}[H]
	\caption{Plot of renewable energy consumption vs. posterior median of latent health with a least squares regression fit}
	\includegraphics[width=0.85\textwidth, height=0.3\textheight]{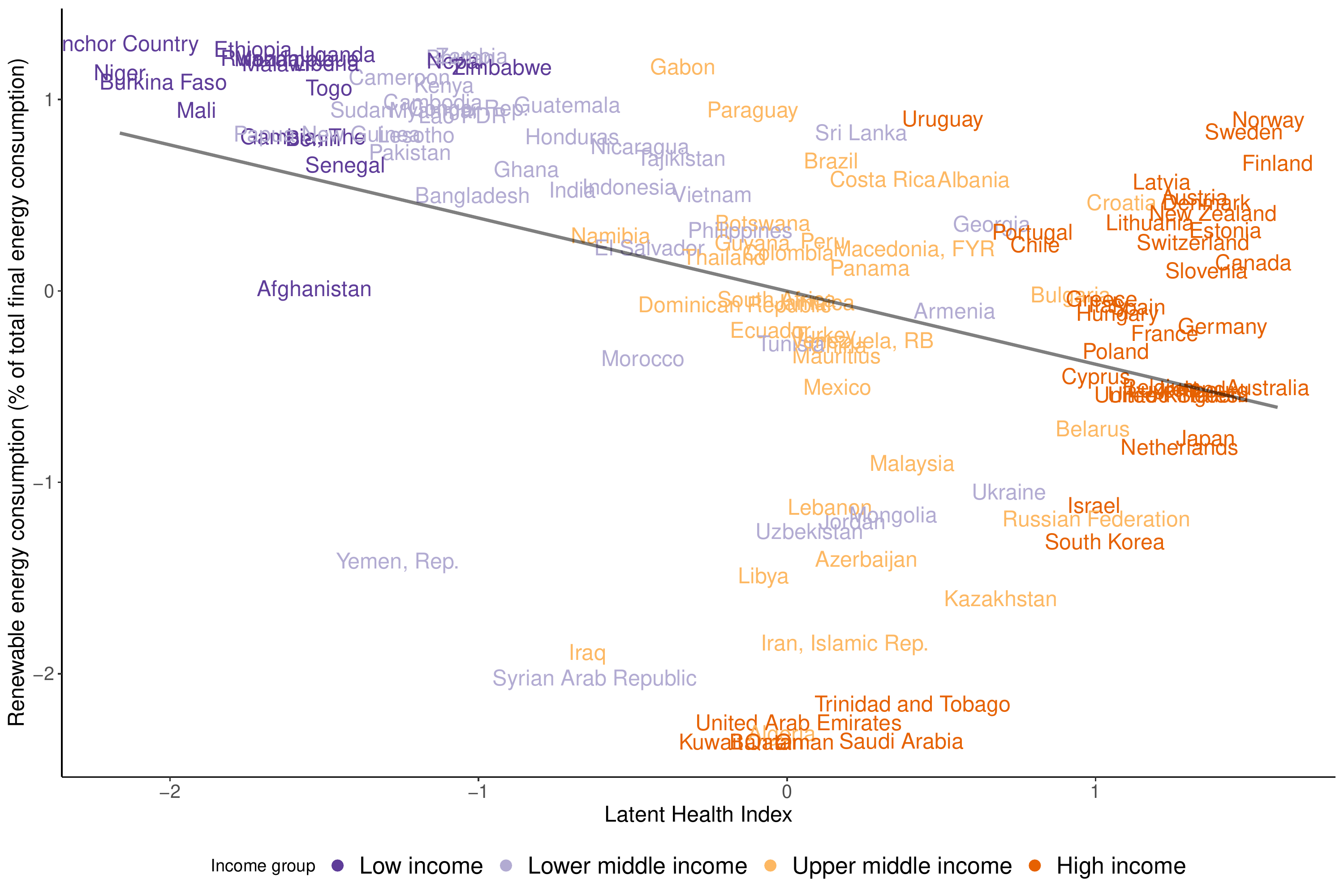}
	\label{fig:renewEnergy}
\end{figure}

\begin{figure}[H]
	\caption{Plot of (reversed-scale) unemployment rate vs. posterior median of latent health with a least squares regression fit}
	\includegraphics[width=0.85\textwidth, height=0.3\textheight]{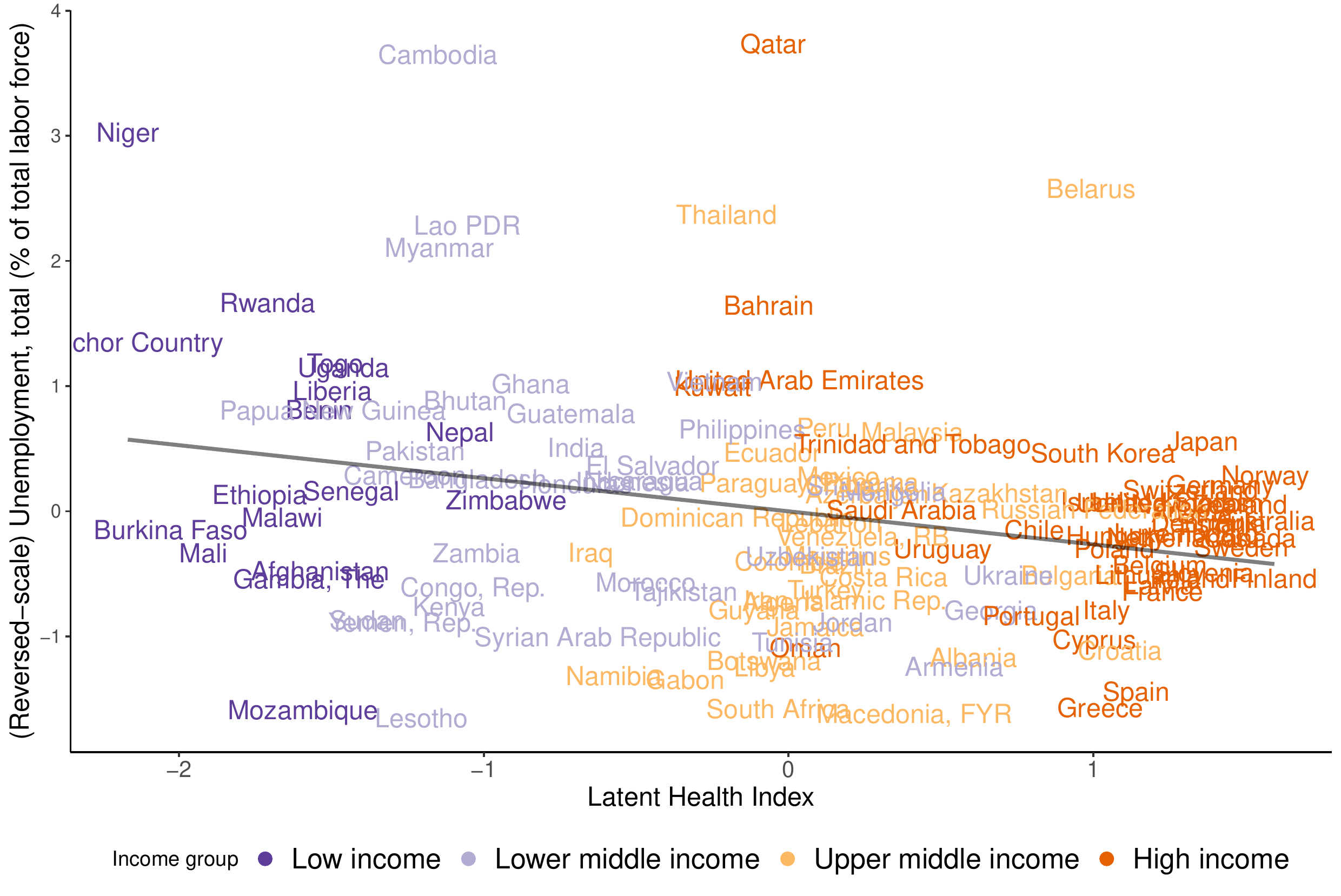}
	\label{fig:unemploy}
\end{figure}

Besides the results discussed in section \ref{ssec:metric}, it was also found that a country's proportion of renewable energy consumption (shown in Figure \ref{fig:renewEnergy}) and the reversed-scale  unemployment rate (shown in Figure \ref{fig:unemploy} and discussed in section \ref{ssec:missing}) each has a negative relationship with the country's latent health (95\% credible intervals for $a_{11} \text{ and } a_{13}$ are $(-0.75, -0.14)$ and $(-0.57, -0.06)$ respectively). The metrics `\texttt{population density}' and `\texttt{proportion of seats held by women in national parliaments}' were found to have no relationship with the latent health of country, conditioned on all the other variables that are in the model. 
 
 \vspace{-1.4em}
\vspace{-2em}

\bibliographystyle{imsart-nameyear}
\bibliography{bibliography}{}

\end{document}